\RequirePackage{fix-cm}
\documentclass[onecolumn]{svjour3}
\smartqed  

\usepackage[square,sort,comma,numbers]{natbib}
\usepackage{booktabs} 
\usepackage[table]{xcolor}
\usepackage[flushleft]{threeparttable}
\usepackage{xcolor}
\usepackage[utf8]{inputenc}
\usepackage{float}
\usepackage{ifthen}
\usepackage{multirow}
\usepackage{caption}
\usepackage{listings}
\usepackage{hyperref}
\hypersetup{colorlinks=true}
\usepackage{url,moreverb,xspace}
\usepackage{tcolorbox}
\usepackage{enumitem}
\usepackage{array,graphicx}
\usepackage{soul}
\usepackage{balance}
\usepackage{pifont}
\usepackage{nicefrac}
\usepackage{mathtools}
\usepackage{amsmath,amssymb,amsfonts}
\usepackage{mathrsfs}
\usepackage{rotating}
\usepackage{ulem}  

\usepackage{csquotes}
\usepackage{adjustbox}

\usepackage{algorithm}
\usepackage{algpseudocode}

\usepackage{tikz}
\usepackage{pgfplots}
\usepgfplotslibrary{statistics}
\usetikzlibrary{fadings}

\usepackage{xcolor,pifont}
\newcommand*\colourcheck[1]{%
  \expandafter\newcommand\csname #1check\endcsname{\textcolor{#1}{\ding{52}}\xspace}%
}
\newcommand*\colourcross[1]{%
  \expandafter\newcommand\csname #1cross\endcsname{\textcolor{#1}{\ding{56}}\xspace}%
}

\newboolean{showcomments}
\setboolean{showcomments}{true}
\ifthenelse{\boolean{showcomments}}
{
	\definecolor{myyellow}{RGB}{255, 228, 26}
	\definecolor{myblue}{RGB}{50, 50, 220}
	\newcommand{\nb}[2]{
		{\sf
			\fcolorbox{myyellow}{yellow}{\scriptsize\textbf{#1}}%
			$\blacktriangleright$%
			{\color{myblue}\fontsize{7pt}{8pt}\selectfont\textbf{#2}}%
		}%
	}
}
{
	\newcommand{\nb}[2]{}
}

\newcommand{\id}{ID}
\newcommand{\xpath}{XPath}

\newcommand{\idxpath}{\id-\xpath{}}
\newcommand{\similo}{Similo\xspace} %
\newcommand{\vonsimilo}{VON Similo\xspace} %
\newcommand{\llmvonsimilo}{LLM VON Similo\xspace} %
\newcommand{\similoplusplus}{Similo++\xspace} %
\newcommand{\vonsimiloplusplus}{VON Similo++\xspace} %
\newcommand{\hybridsimilo}{HybridSimilo\xspace} %

\newcommand{\head}[1]{\noindent\textbf{#1.}}

\newcommand{\comparedpapersimilo}{Nass et al.~\cite{similo}\xspace}
\newcommand{\comparedpapervonsimilo}{Nass et al.~\cite{vonsimilo}\xspace}

\makeatletter
\let\orgdescriptionlabel\descriptionlabel
\renewcommand*{\descriptionlabel}[1]{%
  \let\orglabel\label
  \let\label\@gobble
  \phantomsection
  \edef\@currentlabel{#1}%
  \let\label\orglabel
  \orgdescriptionlabel{#1}%
}
\makeatother

\journalname{EMSE}

\begin{document}

\title{
Web Element Relocalization in Evolving Web Applications: A Comparative Analysis and Extension Study
}

\author{Anton Kluge \and Andrea Stocco
}

\authorrunning{Kluge A. and Stocco A.}
\titlerunning{Web Element Relocalization in Evolving Web Applications}

\institute{A. Kluge \at
	Technical University of Munich -- Boltzmannstra{\ss}e 3 Garching near Munich, Germany \\
	\email{anton.kluge@tum.de} \\ 
        A. Stocco \at 
	Technical University of Munich -- Boltzmannstra{\ss}e 3 Garching near Munich, Germany and fortiss GmbH -- Guerickestra{\ss}e 25 Munich, Germany \\
	\email{andrea.stocco@tum.de|stocco@fortiss.org}
 }

\maketitle

\begin{abstract}
Fragile web tests, primarily caused by locator breakages, are a persistent challenge in web development. Hence, researchers have proposed techniques for web-element re-identification in which algorithms utilize a range of element properties to relocate elements on updated versions of websites based on similarity scoring.
In this paper, we replicate the original studies of the most recent propositions in the literature, namely the Similo algorithm and its successor, VON Similo. We also acknowledge and reconsider assumptions related to threats to validity in the original studies, which prompted additional analysis and the development of mitigation techniques.
Our analysis revealed that VON Similo, despite its novel approach, tends to produce more false positives than Similo. We mitigated these issues through algorithmic refinements and optimization algorithms that enhance parameters and comparison methods across all Similo variants, improving the accuracy of Similo on its original benchmark by 5.62\%.
Moreover, we extend the replicated studies by proposing a larger evaluation benchmark (23$\times$ bigger than the original study) as well as a novel approach that combines the strengths of both Similo and VON Similo, called HybridSimilo. The combined approach achieved a gain comparable to the improved Similo alone. Results on the extended benchmark show that HybridSimilo locates 98.8\% of elements with broken locators in \textcolor{black}{our} testing scenarios.
\end{abstract}


\section{Introduction}\label{sec:introduction}



Automated end-to-end (E2E) web tests created with tools such as Selenium are renowned for being fragile as the the web application under test evolves~\cite{2016-Leotta-JSEP,10132210}. 
Researchers have singled out web element \textit{locators} as the main cause of fragility~\cite{Hammoudi-2016-ICST,Stocco-2018-FSE}. Locators are commands used by test automation tools to identify elements on a web page, hanging on specific properties found in the Document Object Model (DOM), such as the element's identifier, XPath, or text. 


Locator breakages are mainly caused by code changes in the web application. Given the short release cycles and advancements in modern websites, such breakages occur frequently. This poses a significant challenge for automated testing, as it requires manual fixing of tests before they can be executed again. This process is time-consuming and frustrating for testers. As a result, test suites are often abandoned~\cite{6976080}, as the effort outweighs the benefits. 

To mitigate these issues and minimize the number of fragile tests, researchers have proposed automated algorithms that produce robust locators~\cite{2015-leotta-ICST,montoto2010automated,2016-Leotta-JSEP,similo,llmvonsimilo,vonsimilo,color}.
The current state-of-the-art approach for locating a web element corresponding to a specific locator is the Similo algorithm~\cite{similo}. This algorithm calculates a similarity score based on multiple properties of web elements to re-identify the target element among a set of candidates, which consist of all elements on an updated version of the website. The algorithm selects the candidate with the highest similarity score as the new target element. The original algorithm has been extended in two follow-up works, namely VON Similo~\cite{vonsimilo} and LLM VON Similo~\cite{llmvonsimilo}. The former extends the original Similo algorithm by comparing groups of visually overlapping elements on a website instead of singular elements. The latter leverages a large language model to improve the algorithm's accuracy. 


We chose to focus on the Similo algorithm~\cite{similo} and its extensions because recent systematic literature reviews on web application testing~\cite{BALSAM2025112186} identify Similo as the current state-of-the-art approach for repairing broken web test cases. Similo has shown superior performance compared to baseline methods such as Robula+, Vista, and WATER. Additionally, locator-based techniques (e.g., Selenium) have been reported to outperform purely visual techniques in terms of robustness and reliability~\cite{selenium-better-than-visual}. The Similo algorithm is conceptually similar to other locator-identification algorithms like COLOR~\cite{color}, which have demonstrated viability. Moreover, VON Similo and LLM VON Similo, both extensions of Similo, offer novel approaches (e.g., visually overlapping nodes, integration of language models) which motivated us to replicate and further investigate their effectiveness. During our initial review of Similo and its extensions, we observed several limitations and threats to validity in the original evaluations. Initial theories on how to address these limitations led us to extensively re-assess and extend the original results in order to ensure their robustness and applicability. 


First, the original Similo algorithm relies on a fixed set of web element properties and weights to web element re-localization. 
Second, the benchmark used to evaluate Similo contains web elements from versions 12 to 60 months apart. This range does not accurately reflect the actual update frequency of websites, nor it aligns with continuous integration environments, where updates tend to be smaller, more regular, and tests are conducted frequently.
Third, we found a significant revision of the evaluation benchmark between the original study and its subsequent extensions, rather than also assessing the extensions using the initial benchmark for a consistent comparison.

Motivated by the will to understand the causes of these discrepancies and to address the aforementioned challenges, in this paper, we replicate the \similo~\cite{similo} and \vonsimilo~\cite{vonsimilo} studies, improving the experimental setting of the original papers to address the identified limitations and threats to the validity. More in detail: (1)~we improve Similo by optimizing the attributes and weights of the original algorithm. We evaluated six \textit{new} similarity functions to calculate the similarity between web element properties and optimize the weights assigned to these similarities using a genetic algorithm. We also analyze the capabilities of a novel hybrid version that combines the capabilities of Similo and VON Similo. (2)~We collected a benchmark dataset of more than 10,000 element pairs from multiple websites and versions over the past five years. Our benchmark is 12$\times$ bigger than the original Similo benchmark and 23$\times$ bigger than the \vonsimilo benchmark, and it is more reflective of the fine-grained modifications occurring in real web sites. (3)~We perform a fairer comparison between the original Similo and its extensions by evaluating all algorithms on the same benchmarks, both using the original as well as our new extended benchmark. 
While we were able to replicate the original study results, our findings are in contrast with ones by \comparedpapervonsimilo as we show that VON Similo under-performs Similo in directly identifying the target element but excels in identifying the visual overlap of the target element. 

Our paper makes the following contributions:

\begin{itemize}[noitemsep]
    \item \textbf{Replication.} A replication study of the results of the Similo and VON Similo algorithms, including a comparison with the experimental setup and benchmarks used in the Similo and LLM VON Similo algorithms. Ours is the first attempt at evaluating all Similo and VON Similo algorithms on the same benchmark under analogous conditions.
    \item \textbf{Extended Benchmark and Metrics.} We extended the original benchmark of 804 element pairs to 10,376 element pairs and introduced six additional metrics.
    \item \textbf{Library.} A library for the Selenium framework, which is publicly available~\cite{replication-package}. Our tool wraps an existing locator and uses our extended \similo algorithm to locate an element if the original locator fails. 
\end{itemize}


\section{Motivating Example}\label{sec:background}

\begin{figure*}[t]
\centering
\includegraphics[scale=0.45]{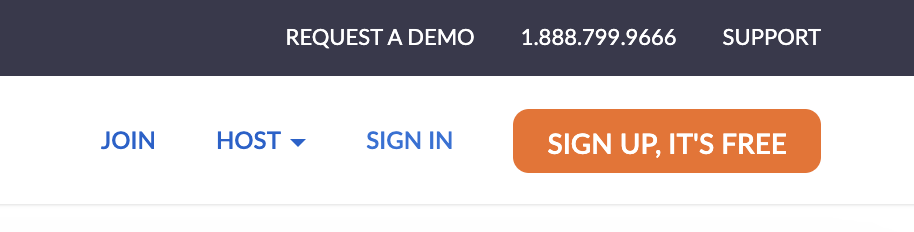}
\label{fig:zoom_old_example}
\hfill
\includegraphics[scale=0.42]{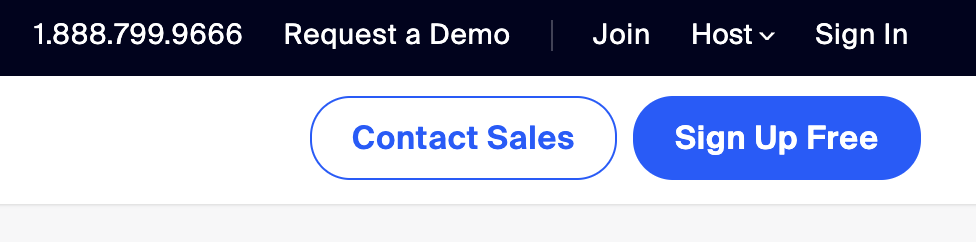} %
\label{fig:zoom_new_example} 
\caption{Zoom.us homepage on September 2022 (top) and January 2023 (bottom).} 
\label{fig:zoom_example} 
\end{figure*}

In this section, we describe the problems occurring to E2E web tests during web app evolution.  
We use as a running example the home page of Zoom.us, a popular online chat service used for video communications, messaging, voice calls, conference rooms for video meetings, and virtual events. 
\autoref{fig:zoom_example}~(top) shows the website on September 2022, whereas \autoref{fig:zoom_example}~(bottom) shows the update website on January 2023. 
In the short timespan of four months, the website has undergone a significant redesign, which has affected the locators of the elements and likely broke possible test cases.
In addition to stylistic changes, some buttons and links were relocated in the GUI (e.g., the Host drop-down list). In the rest of this section, we describe the variety of breakage scenarios that can occur in web tests, which robust relocalization techniques should aim to handle~\cite{Stocco-2018-FSE}.


\subsection{Element Not Found}
When the provided locator fails to identify an element on an updated website, the test case will break~\cite{Stocco-2018-FSE,Hammoudi-2016-ICST}. This requires a developer to manually locate the element on the updated website and modify the locator in the test case.

Automated techniques such as Similo aim to correctly identify the element on the updated version of the website. The algorithm requires a working reference web app, typically an old version of the website in which the web test used to function, to gather a set of properties. These properties are then used to try find the element on the new version of the website. For instance, the ``Request a Demo'' button has the same tag and non-capitalized text, and a similar shape, location, and neighboring text. All such information are used by Similo in correctly identifying the link on the updated website. By comparing the properties of the old element with the properties of all elements on the new website, Similo returns the most similar element, which likely identifies the original web element. 

\subsection{False Positive} Another common reason for test breakages occurs when the locator returns another existing element instead of the intended target~\cite{Stocco-2018-FSE}. Repairing these breakages is particularly challenging because the developer has to manually trace where the test deviates from its intended path. For instance, the button initiating a direct call to 1.888.799.9666 in \autoref{fig:zoom_example}~(top) can be identified by the CSS Selector \texttt{\#black-topbar > div > ul > li:nth-child(2) > a}~\cite{MDNCSSSelectors}. The same locator will return the ``Support'' button in \autoref{fig:zoom_example}~(bottom), even though they serve different purposes. A test case will not break immediately, as the locator returns an existing element, but later on after the test case execution has deviated from the intended path. \similo can help identify these false positives by validating if another element on the updated website. 

In case of correct detection, \similo could actually support the automated repair of the broken locators. While \similo does not store the information required to locate the element in the source code, as the amount of data collected would clutter the test case, it could in practice use some form of caching mechanism to store the information from the old version of the website. By saving an identifiers for each web element in cache, \similo could update the information associated with such web elements on new evolved DOM versions.

\subsection{Misclassifications} 
The internal functioning of \similo may cause the algorithm to misclassify elements. One reason is due to the target element changing its tag. As the tag property is highly weighted in \similo calculations, other elements in the close proximity of the target which have the same tag as the target element can be misclassified. For example, the ``Join'' button in \autoref{fig:zoom_example}~(top) has the tag \texttt{a}, and the tag \texttt{button} in \autoref{fig:zoom_example}~(bottom). Because the ``Sign Up Free'' button in \autoref{fig:zoom_example}~(bottom) has the same tag as the target element, as well as similar shape, location, neighboring text and XPath, \similo misclassifies the ``Sign Up Free'' button as the ``Join'' button.
This is especially problematic as \similo always returns an element. If the element is not the intended target, the test case will break at an arbitrary point in the test execution, making it harder for developers to identify the root cause of the breakage.

\section{Approaches}\label{sec:approaches}

This paper is a replication and extension of the work by \comparedpapersimilo presented in the ACM Transactions on Software Engineering and Methodology (vol. 32, no. 3) in 2023.

The algorithm is based on the idea that when a web element is modified, some properties are altered while others remain the same or undergo minor changes. Thus, the main working assumption is that by calculating a similarity score between two elements, the algorithm can identify the element with the highest similarity score as the target element. Two successors to the basic Similo algorithm have been developed: \vonsimilo~\cite{vonsimilo} and LLM \vonsimilo~\cite{llmvonsimilo}. The first successor, \vonsimilo, takes advantage of the fact that visual entities on a website often consist of multiple concrete web elements, which can help to identify the target element. The second successor (LLM Von Similo) utilizes a Large Language Model (LLM) to locate the correct element from a pre-selected set of candidates.

For all Similo variants, the algorithm starts with an element that can be reliably identified on an old, baseline version of a specific website. This element is referred to as the target. The algorithm extracts all necessary information about this element. Then, the algorithm tries to find the target element among all candidates elements on a new version of the same website, where the locator used to identify the target element on the baseline version no longer works. 
The algorithm calculates a similarity score between the target element and all candidates. The candidate with the highest similarity score is then returned as the target element. It is important to note that \similo is not a locator algorithm. It does not generate a robust locator for the target element, but rather identifies the target element among all candidates. A robust locator for the target element must be generated using other methods, such as Robula+~\cite{2016-Leotta-JSEP}, COLOR~\cite{color}, or the MultiLocator~\cite{2015-leotta-ICST}.

%
In the following sections, we describe each algorithms in a greater level of detail, using the following nomenclature. 
$E$ will always refer to an arbitrary web element. $T$ will be used for target elements, the original version of an element that needs to be found on an updated website, and $C$ for a possible candidate for the target on the updated website. The candidate on the new version of a website that corresponds to the target element will be noted as $C'$. If multiple elements are present, they will be referred to as $E_1, .., E_n$. When working with properties, $E_n.a_{m}$ refers to the $m$-th property of the $n$-th element. A visual overlap for the element $E$ is noted as $O^E$.

\subsection{\similo}\label{sec:similo}


\begin{algorithm}[t]
\caption{Similo}
\scriptsize
\label{alg:similo}
\begin{algorithmic}[1]

\Function{Similo}{$target\_element, candidate\_elements$}
    \State $best\_score \gets 0$
    \State $best\_candidate \gets \text{null}$

    \ForAll{$candidate \in candidate\_elements$}
        \State $score \gets$ \Call{SimiloScore}{$target\_element, candidate$}
        \If{$score > best\_score$}
            \State $best\_score \gets score$
            \State $best\_candidate \gets candidate$
        \EndIf
    \EndFor

    \State \Return $best\_candidate$
\EndFunction

\vspace{0.5em}

\Function{SimiloScore}{$target, candidate$}
    \State $total \gets 0$

    \ForAll{$property \in target.properties$}
        \State $t \gets$ \Call{Get}{$target, property$}
        \State $c \gets$ \Call{Get}{$candidate, property$}
        \State $sim \gets$ \Call{Similarity}{$t, c$}
        \State $total \gets total + sim \times WEIGHTS[property]$
    \EndFor

    \State \Return $total$
\EndFunction

\end{algorithmic}
\end{algorithm}

Similo (Algorithm~\ref{alg:similo}) requires a working version of the web application where the web element can be identified. Given the original web element $T$ and its properties $T.a_1..., T.a_n$, as well as all new web elements $C_1, ..., C_n$, of which one should be the original one $C'$, Similo computes a similarity score between $T$ and each $C_1, ..., C_n$ individually:

$$\text{Similo}(T, C) = \sum_{i \in \#\text{properties}} \text{similarity}(C.a_i, T.a_i) \cdot c_i$$ 

Here, $c_i$ is a property-specific weight based on the COLOR study~\cite{color}, determining how effective, stable, and unique specific properties are. Stable properties, such as tag and name, are assigned a weight of 1.5, while non-stable properties receive a weight of 0.5. The similarity function calculates a score ranging from $[0, 1]$ on how similar both values are using a predetermined algorithm. This score is used to generate a ranking of all web elements based on their similarity with the original element, enabling developers to choose the element with the highest similarity score or retain all elements above a certain threshold.

The specific properties used in the Similo algorithm can be found in \autoref{tab:properties_similo}. These properties are selected based on the locator types supported by the Selenium WebDriver API (id, name, class, tag, link text, partial link text, \xpath{}, and CSS) for native element location. The authors also consider the locators chosen by Selenium idE (a tool for recording and replaying user interactions), including id, link text, name, and various XPaths. Additional properties are chosen based on the results of the COLOR study (id, class, name, value, type, tag name, alt, src, href, size, onclick, height, width, \xpath{}, X-axis, Y-axis, link text, label, and image) and the WATER study~\cite{water} (id, \xpath{}, class, link text, name, tag, coord, clickable, visible, z-index, hash). The authors selected all DOM-based properties from this list and excluded properties extracted from the visual user interface. Properties with slight differences between versions, such as class, links, and XPaths, are compared using the Levenshtein distance. On the other hand, properties that tend to change completely, such as tag and id, are compared using equality. Integer-based properties are compared using the Euclidean distance. 

\begin{table}[t]
    \caption{Properties, similarity functions, and weights used by Similo~\cite{similo}.}
    \centering
    \footnotesize

    \setlength{\tabcolsep}{9.2pt}
    \renewcommand{\arraystretch}{1}

    \begin{tabular}{@{}llcl@{}}
    \toprule
    \textbf{} & \textbf{Similarity Function} & \textbf{Weight} & \textbf{Description} \\
    \midrule
    Tag & Equality & 1.5 & Element type \\
    Class & Levenshtein Similarity & 0.5 & CSS class \\
    Name & Equality & 1.5 & Attribute value \\
    ID & Equality & 1.5 & Unique identifier \\
    HRef & Levenshtein Similarity & 0.5 & Link address \\
    Alt & Levenshtein Similarity & 0.5 & Image text \\
    Absolute \xpath{} & Levenshtein Similarity & 0.5 & Full DOM path \\
    \idxpath{} & Levenshtein Similarity & 0.5 & Relative DOM path \\
    Is Button & Equality & 0.5 & \texttt{button} tag or \texttt{a} tag with \texttt{btn} \\  & & & in class or a \texttt{input} with button, \\ & & & sumbit or reset type. \\ 
    Location & Euclidean Distance & 0.5 &  X and Y coordinates \\
    Area & Euclidean Distance & 0.5 & Calculated as width * height \\
    Shape & Euclidean Distance & 0.5 & Calculated as width / height \\
    Visible Text & Levenshtein Distance & 1.5 & Value, inner text or placeholder \\
    Neighbor Texts & Word set similarity & 1.5 & All visible text in a larger \\ & & & rectangle around the element \\
    \bottomrule
    \end{tabular}
    \label{tab:properties_similo}
\end{table}

\subsection{\vonsimilo}\label{sec:vonsimilo}

In this paper we also replicate and extend \vonsimilo, an extension of the basic \similo algorithm introduced by \comparedpapervonsimilo and presented in the IEEE Conference on Software Testing, Verification and Validation (ICST) in 2023~\cite{vonsimilo}.


\begin{algorithm}[H]
\caption{Von Similo}
\scriptsize
\label{alg:von-similo}
\begin{algorithmic}[1]

\Require original element $T$, old web page $OldSite$, candidate elements $C$
\Ensure best matching element in $C$

\Function{VonSimilo}{$T, OldSite, C$}
    \State $T\_overlap \gets$ \Call{Overlap}{$T, OldSite$}
    \State \Call{UpdateProperties}{$T, T\_overlap$}
    \State $best\_score \gets 0$
    \State $best\_candidate \gets \text{null}$
    \ForAll{$cand \in C$}
        \State $cand\_overlap \gets$ \Call{Overlap}{$cand, C$}
        \State \Call{UpdateProperties}{$cand, cand\_overlap$}
        \State $score \gets$ \Call{VonSimiloScore}{$T, cand$}
        \If{$score > best\_score$}
            \State $best\_score \gets score$
            \State $best\_candidate \gets cand$
        \EndIf
    \EndFor
    \State \Return $best\_candidate$
\EndFunction

\vspace{0.75em}
\Function{VonSimiloScore}{$T, cand$}
    \State $total \gets 0$
    \ForAll{property lists $(L_T, L_c)$ in \Call{Zip}{$T.properties, cand.properties$}}
        \State $m \gets 0$
        \ForAll{$p_T \in L_T$}
            \ForAll{$p_c \in L_c$}
                \State $s \gets$ \Call{Similarity}{$p_T, p_c$}
                \If{$s > m$}
                    \State $m \gets s$
                \EndIf
            \EndFor
        \EndFor
        \State $total \gets total + m$
    \EndFor
    \State \Return $total$
\EndFunction

\vspace{0.75em}
\Function{UpdateProperties}{$e, OverlapSet$}
    \ForAll{$prop \in e.properties$}
        \State $values \gets [\ ]$
        \ForAll{$o \in OverlapSet$}
            \State append $o.properties[prop.name].value$ to $values$
        \EndFor
        \State $prop.value \gets values$
    \EndFor
\EndFunction

\vspace{0.75em}
\Function{Overlap}{$e, S$}
    \State $O \gets [\ ]$
    \ForAll{$cand \in S$}
        \If{\Call{Overlaps}{$e, cand$}}
            \State append $cand$ to $O$
        \EndIf
    \EndFor
    \State \Return $O$
\EndFunction

\vspace{0.75em}
\Function{Overlaps}{$e, cand$}
    \State $r_T \gets e.rect()$
    \State $r_C \gets cand.rect()$
    \State $ratio \gets \Call{IntersectArea}{r_T, r_C} \,/\, \Call{UnionArea}{r_T, r_C}$
    \State $inside \gets \Call{IsInside}{cand.center(), r_T}$
    \State \Return $(inside \wedge ratio \ge 0.85)$
\EndFunction

\end{algorithmic}
\end{algorithm}

The idea behind \vonsimilo (Algorithm~\ref{alg:von-similo}) is that web elements often consist of multiple parts. The DOM orders its elements so that a node's children are inside the area the parent occupies on the visually rendered website. They appear as one visual unit to the user and can be interacted with as one unit. 

For example, a button might contain a button tag, an icon, and text. It does not matter which exact part the user clicks; the button will be triggered. When two elements have a considerable overlap, meaning they share a large part of their occupied area, they are likely to be part of the same visual unit. This unit is called visually overlapping nodes, or VON.

When a web element is moved to a different location on the website, the nodes in its visual overlap are often moved as well. The combination of elements moved together provides a more unique fingerprint than a single web element would. The situation is similar when a node is changed; its visual overlap might stay the same, making it easier to identify the modified element by the combination of elements in its visual overlap. For example, the button in our example moves to a different location on the website, but the icon and text stay the same. Alternatively, the text inside the button changes, but the button and icon stay the same.

The improved algorithm in \vonsimilo leverages this heuristic to identify elements more reliably. When a target element needs to be found on a new website version, the algorithm first identifies the visual overlap of the target element on the baseline version. Then, it iterates over all candidates on the updated version and calculates their respective visual overlaps. In the last step, a score is calculated between the target visual overlap of nodes and all potential candidate visual overlaps. The score is calculated between two elements, but includes all the property values of the visually overlapping nodes. For each property all possible combinations of values are compared and the maximum similarity is weighted and added to the total score. Similar to \similo, the candidate with the highest score is returned as the target element in the new version.

To calculate the score, we need to define the visual overlap of a node. The VON Similo paper defines that two web elements $E_1$ and $E_2$ are visually overlapping if the following two conditions are met:

\begin{enumerate}
	\item The areas $R_1$ and $R_2$, which their respective rectangles occupy on the screen in pixels, intersect to a certain degree. In other words if $\frac{R_1 \cap R_2}{R_1 \cup R_2}$ is above a certain threshold, which should be chosen in a way that balances:
		\begin{enumerate}
			\item accidentally grouping elements that do not belong together, when the threshold is chosen to lose, and there is no significant overlap, and
			\item not recognizing two visually overlapping nodes as such by choosing the threshold too high.  
		\end{enumerate}
	The authors propose a threshold of $0.85$.
	
	\item The center of $E_2$ is located inside of $R_1$. (The paper describes this differently - \enquote{The center of the web element $W_1$ (here $E_1$) is contained in the rectangle $R_1$} - but the underlying code shows that $E_2$ is compared with $R_1$).
\end{enumerate}

In addition the LLM VON Similo paper introduces another metric to determine if $E_1$ and $E_2$ are visually overlapping based of the nodes visual text and XPath, which is used in addition to the algorithm used in VON Similo. We will called this approach textual overlap, and it is defined as:

\begin{enumerate}
    \item The visible text of both elements is not \texttt{null} and their content is case-sensitive equal.
    \item The absolute XPath of $E_1$ is a prefix of the absolute XPath of $E_2$, e.g, $E_2$ is a child of $E_2$ in the DOM.
\end{enumerate}

After calculating all overlapping elements $E_1, ..., E_n$ for one element $E$, we replace the properties $E.a_1, ..., E.a_m$ of that element with lists of property values of the overlapping nodes, meaning the properties of $E$ would look like this: $[E_1.a_1, ..., E_n.a_1],\allowbreak ...,\allowbreak [E_1.a_m, ..., E_n.a_m]$. 

To compare two elements $T$ and $C$, whose properties have been replaced with their respective lists, we modify the Similo calculation by choosing the pair from both property lists with the highest similarity. We then take the sum of those maximized values:

$$ \text{VONSimilo}(T,C) = \sum_{i \in \#\text{properties}} \Big(\max_{t \in T.a_i, c \in C.a_i} \text{similarity}(t, c)\Big) \cdot c_i$$ 

\subsection{LLM \vonsimilo }\label{sec:llmvonsimilo}

\llmvonsimilo is the latest iteration of the \similo algorithm~\cite{llmvonsimilo}. 
The algorithm first uses \vonsimilo to rank all elements on the website by their score. It then takes the top 10 visual overlaps, as the target element is most likely among them. The algorithm then utilizes GPT-4~\cite{gpt4} to find the target element among the pre-selected candidates. The algorithm converts the ten elements and the target element into JSON format. It then sends a request to the large language model with the target and the ten candidates, asking it to identify the target's match. The LLM responds with a single number, indicating the target's position in the list of candidate elements.

Since \llmvonsimilo relies on proprietary OpenAI's APIs and uses an unspecified version of GPT-4 with unknown temperature and configuration settings, its results cannot be reliably reproduced in a controlled experimental environment. Therefore, we do not attempt a full replication of this method in this paper. 
Rather, in our replication, we only consider the benchmark used in \llmvonsimilo and the \vonsimilo implementation and test the other algorithms against it.


\subsection{Limitations and Threats to Validity}\label{sec:shortcomings}

In this section we describe the limitations we identified about the the Similo algorithms, as well as the threats to the validity of the original studies that we aim to address in our replication work.  



\head{Threat T1 $\cdot$ Changed benchmarks}
The benchmarks and underlying metrics vary across all three versions of Similo, making it challenging to accurately compare and evaluate each algorithm. We could identify three major inconsistencies between benchmarks, used metrics and evaluation strategies:

\textit{Changed data.} 
The websites and elements used for the evaluation in the different benchmarks differs between the papers. The Similo and LLM \vonsimilo have approximately the same data (48 websites, $\approx$800 element pairs) with minor differences. The \vonsimilo paper only uses 36 websites and around 400 element pairs.
    
\textit{Changed metric.} The Similo and LLM VON Similo benchmarks use a setup where the Similo algorithm is used to rank all elements on an updated version of a website by their similarity with the target, choosing the highest ranking one and validating whether it is the actual target. We used the same setup for all subsequent benchmarks. The VON Similo paper, on the other hand, calculates the similarity score between the target on the new and old version, declaring it a match when if the score reaches a certain threshold. We believe that this setup is not reflecting how Similo will perform in a web testing scenario. Additionally, it does not allow us to properly compare Similo with VON or LLM VON Similo as they have both never executed on the benchmark.

\textit{Changed similarity functions.} The used similarity functions change slightly between the different iterations of Similo. For example, the original Similo algorithm uses Levenshtein similarity on the raw text, while the others apply it to lower cased strings. 

\head{Threat T2 $\cdot$ Coarse granularity of version snapshots}
The benchmark used for all Similo papers utilizes website versions that were collected between 12 and 60 months apart. This scenario does not reflect actual web testing practices, where tests are run in a regular schedule, e.g., nightly or over the weekend. The changes made over a 12-60 month period are rarely made between two consecutive test runs of a test~\cite{Hammoudi-2016-FSE}. Hence, Similo can use the smaller update steps to repair itself by updating the saved value for a certain element. 
Utilizing the current benchmarks presents an unrealistic perspective on \similo's effectiveness in a more realistic testing time frame, thereby underestimating its true capabilities.


\head{Threat T3 $\cdot$ Fixed set of properties and weights}
The properties, similarity functions and weights used in the Similo algorithms were taken from the COLOR study~\cite{color} or given without empirical evidence that these values are suitable to be used in such an algorithm. Furthermore, they are reused for the VON Similo algorithm which works differently than the Similo algorithm. This reuse was implemented without assessing whether alternative properties, similarity functions, and weights might better suit the VON Similo's distinct computational approach, which emphasizes visual overlap.

\head{Threat T4 $\cdot$ Limited accuracy with VON Similo}
Algorithms that use visual overlap can identify only the group of elements that overlap visually, not the specific target within that group. For certain test actions, such as clicking a button or link, this does not represent an issue because the website interprets any click within the overlapping area as a click on the element itself. However, for other actions like entering text into an input field, a text area, or verifying specific properties of a particular element, merely interacting with any element in the overlapping area is insufficient. Instead, identification of the exact element is necessary.

The underlying problem is that given $T$ and $C'$ as well as elements $E^T_{1..n}$ and $E^{C'}_{1..m}$ which form their respective visual overlaps $O^T$ and $O^{C'}$, then 
this visual overlap is the same for all elements in the overlap, e.g. $\forall C \in O^{C'}: \texttt{overlap}(C) \equiv O^{C'}$. When Similo compares $T$, with every $E^{C'}_{1..m}$, it will always compare their visual overlaps: $O^T$ and $O^{C'}$ and calculate the same score each time. Because the algorithm chooses the element with the highest score, it selects a random element from the visual overlap $O^{C'}$.

\subsection{Implementation}



To be able to \textcolor{black}{evaluate} the Similo algorithm in a practical setting, we implemented a library for Java, compatible with the Selenium WebDriver framework. The library wraps standard Selenium locators and automatically uses the Similo algorithm to identify the correct web element when the original locator fails. Upon first usage or locator changes, the library captures relevant element properties and saves them in an SQL database to do accurate matching in subsequent test executions. The wrapper supports all Selenium's locator strategies (e.g., XPath, CSS selectors, ID), minimizing integration effort with existing test suites.

\textcolor{black}{The implementation features built-in self-repair capabilities by locating the element whose locator is broken and using it to continue test execution. It also intelligently updates locators within the database when elements change by determining the most stable of the basic locators \xpath{}, \id{} and \idxpath{} and using it for future localizations.} Doing so, it enhances test resilience without modifying test code directly but remembering the connection between the locator and its current state of properties. Moreover, the library monitors Similo scores and issues configurable warnings when low score matches occur, guiding developers toward potentially broken locators. The library is available on GitHub and has comprehensive documentation~\cite{replication-package}.

\section{Empirical Study}\label{sec:study}

\subsection{Research Questions}

We consider the following research questions:

\noindent
\textbf{RQ\textsubscript{0} (replication):} Can we replicate the experimental results yielded by state-of-the-art tools targeting robust locator generation? 

\noindent
\textbf{RQ\textsubscript{1} (comparison):} How do Similo, VON Similo, and LLM VON Similo compare with each other on the same benchmark? 

\noindent
\textbf{RQ\textsubscript{2} (improvements):} How do effectiveness vary when considering different metrics in \similo and \vonsimilo? 

\noindent
\textbf{RQ\textsubscript{3} (hybrid):} How do effectiveness vary when combining \similo and \vonsimilo? 

In the first research question (RQ\textsubscript{0}) we aim to confirm the reliability of existing robust locator generation approaches \similo and \vonsimilo by reproducing their experimental results against their original data sets.
The second research questions (RQ\textsubscript{1}) addresses \textbf{T1} by performing a comparison by executing the selected algorithms on four benchmarks, three taken from the original papers, and a new one substantially extended in this work (hence addressing \textbf{T2}).
The third research question (RQ\textsubscript{2}) evaluates a large set of configurations of the original algorithms, varying the metrics being used to optimize the accuracy of the algorithm on different metrics. This question aims to address the problems discussed in \textbf{T3}.
The last research question (RQ\textsubscript{3}) evaluates \hybridsimilo, a novel hybrid approach in which we combine \similo and \vonsimilo. This question aims to address the problems discussed in \textbf{T4}. 

\subsection{Benchmarks}\label{sec:benchmarks}

To mitigate the problems associated with \textbf{T1}, in this paper we benchmark all algorithms on all available benchmark available in the original \similo, \vonsimilo, and \llmvonsimilo papers. Particularly, all benchmarks use the Web Archive~\cite{web-archive} to collect the element pairs and their property values.
We also mitigate \textbf{T2} by providing an extended benchmark constructed using the websites as the original benchmarks and the same selection criteria for web elements, but sampling a higher number of web elements and versions over time. 
All the locators used in the original benchmarks were taken from the corresponding replications packages. The used websites were reloaded from the same WayBack Archive links.

\subsubsection{Similo's Benchmark}\label{sec:similobenchmark} 

The benchmark used to evaluate the original Similo algorithm consists of 809 web element pairs $T$ and $C'$, from 48 websites, each having 12 to 60 months between versions.

\subsubsection{VON Similo's Benchmark}\label{sec:vonsimilobenchmark}

For the VON Similo benchmark 442 web elements from 33 websites were used, each having 12 to 60 months between version, similar to the Similo benchmark. Additionally all the elements from the visual overlaps of the targets were added, resulting in 1,163 element pairs. Finally for every matching element pair, one randomly selected non matching pair was added. The non matching pairs were selected to measure the false positive and false negative rate of the algorithm.
The 442 base web elements of the VON Similo benchmark are a subset of the 809 web elements of the Similo benchmark. The paper does not provide any information why a smaller benchmark was used for VON Similo.

\subsubsection{LLM VON Similo's Benchmark}\label{sec:llmvonsimilobenchmark}

The LLM VON Similo benchmark utilizes the same website versions as Similo and shares 804 element pairs. The other five elements which were part of the original benchmark could not be located due to changed rendering when reloading the websites from the WayBack Archive.

\subsubsection{Extended Benchmark}\label{sec:extendedbenchmark}


We collected web elements from the homepages of 30 popular web applications, considering the 48 websites used in original papers and combining it additional websites from a website ranking from 2023~\cite{similarweb}. We had to discard 18 websites that contained broken snapshots or did not render properly. Five, only partially broken websites were used for later cross-validation of our training approach (further details are available in our replication package). 

To better reflect to short time spans between test executions and to improve comparability, we chose a fixed time span between two version of four months and selected 16 version from September 2018 to September 2023 for each web application. We then tracked elements across those versions, resulting in a total of 933 inital elements from these websites and 10,376 element pairs of the entire time span. This process took three months of manual work for mapping each element across all versions.
We also classified the element pairs in categories, determined if any of the basic unqiue locators (ID, XPath, or ID-XPath) changed or if it was still locatable using the one of these. This information is used to later benchmark the performance of the algorithms on locator pairs with broken unique locators.
The total distribution of elements across the websites is shown in the appendix in~\autoref{tab:classification_distribution}.

\subsection{Algorithms Extensions}\label{sec:extensions}

To mitigate the problems associated with \textbf{T3} and \textbf{T4}, we devised three extensions of the original \similo and \vonsimilo algorithms, introducing a hybrid method that leverages both. 

\subsubsection{\similoplusplus}
\label{sec:similopp}

The first improvement to Similo is called \similoplusplus. In brief, we improve the properties compared by the algorithm, the comparison algorithms used to compute the similarity between two properties and the weights which are multiplied with the similarity.

\head{Compared Properties} 
Along with the properties used in the original Similo algorithm, we analyzed the frequency (how many element pairs have that property) and stability (in how many cases has the property the same value on the updated version). We found two additional suitable attributes: \textcolor{black}{\texttt{type} (frequency of 6-7\% and stability of 95-96\%) and \texttt{aria-label} (frequency of 10-12\% and stability of 81-84\%) across five random subsets of the training data as well as the cross validation set.} Additional we compare all attributes for that element as a key-value map.  


\head{Similarity Functions}
The original paper utilizes simple (lower case) equality, Levensthein distance and word comparison to compare string properties, Euclidian distance for integer fields like area and shape as well as 2D-Distance for the coordinates. We extended these comparison algorithms by consider additional distance metrics. For string based properties, we considered the following additional comparison algorithms:

\textit{Jaccard.}
The Jaccard distance is originally a measure to compare two sets and is defined as the size of the intersection divided by the size of the union of the input sets $A$ and $B$, $\frac{|A \cap B|}{|A \cup B|}$. To use the Jaccard distance to compare strings, we first split the strings into sets of characters and then use the Jaccard distance to compare the sets. The Jaccard distance between \enquote{kitten} and \enquote{sitting} is $\frac{3}{7}$, because the intersection of the sets is \{\enquote{t}, \enquote{i}, \enquote{n}\} and the union is \{\enquote{k}, \enquote{i}, \enquote{t}, \enquote{e}, \enquote{n}, \enquote{s}\}. 

\textit{Jaro Winkler.} 
The Jaro-Winkler distance is a string metric measuring the edit distance between two sequences. The Jaro-Winkler distance is given by a modification of the Jaro distance formula, where more weight is given to strings that match from the beginning. The Jaro distance is given by $d = 1 - \frac{1}{3}(\frac{m}{|s_1|} + \frac{m}{|s_2|} + \frac{m - t}{m})$, where $m$ is the number of matching characters and $t$ is half the number of transpositions. A transposition is a pair of matching characters in the wrong order in one of the strings. The Jaro-Winkler similarity is then calculated as $\text{Jaro similarity} + l \times p \times (1 - \text{Jaro similarity})$, where $l$ is the length of the common prefix at the start of the string up to a maximum of four characters, and $p$ is a constant scaling factor, often 0.1. For example, the Jaro-Winkler similarity between \enquote{kitten} and \enquote{sitting} is approximately 0.74, assuming the Jaro distance is 0.77, and there is no common prefix.

\textit{Set similarity.} 
The set similarity is similar to the Jaccard similarity. The strings are split into sets of strings at spaces and newlines. The final result is the Jaccard distance between the two sets. For example, the similarity between \enquote{Sign up} and \enquote{Sign in} would result in $\frac{1}{3}$, because the intersection is \{\enquote{Sign}\} and the union is \{\enquote{Sign}, \enquote{up}, \enquote{in}\}. Capitalization is ignored.

To compare properties which consist of key value pairs, i.e. the elements attributes, we considered the following two algorithms:

\textit{Intersect Value Compare.} It is calculated as 
    $$\frac{|\{(k, v) | (k, v) \in A \cap B \}|}{\max(|A|,|B|)}$$
    where $A.k$ is the value of the key $k$ in the map $A$.

\textit{Intersect Key Compare.} It is calculated as 
    $$ \frac{|\{k | k \in A \cap B \}|}{|\{k | k \in A \cup B \}|}$$

To compare the distance between two elements we introduced two new similarity algorithms:

\quad\textit{Manhattan Distance.}
The distance between two points in a grid based on a strictly horizontal and/or vertical path. The Manhattan distance between the points $(x_1, y_1)$ and $(x_2, y_2)$ is $|x_1 - x_2| + |y_1 - y_2|$. The result is normalized to $[0, 1]$ by dividing it by a predefined maximum distance.

\quad\textit{Exponential decay.} uses the Euclidean distance but adds exponential decay, with differing $\lambda$ values calculated as $e^{-\lambda d}$, where $d$ is the Euclidean distance between the points. The benefit is that the function is already normalized to $[0, 1]$ and approaches $0$, eliminating the need to define a maximum distance. We used different values for $\lambda$, evaluating three exponential decay similarity functions, namely $\lambda = 0.001$ as a small decay, $\lambda = 0.005$ as a medium, and $\lambda = 0.01$ as a large decay.

Finally to compare the shape of the two elements we used:

\textit{Area.}
The area of an element is the product of the width and height of the minimal rectangle containing the visual element. 

\textit{Perimeter.} 
The perimeter of an element is the sum of the length of all sides of the minimal rectangle containing the visual element.

\textit{Aspect Ratio.} 
The aspect ratio of an element is the ratio of the width to the height of the minimal rectangle containing the visual element.
To compare the respective values of two elements, we divide the smaller value by the larger value. For example, comparing an element with an area of 100px 
and an element with an area of 200px would result in $\frac{1}{2}$ because the smaller element has half the area of the larger element.

\head{Improved Weights}
We use a genetic algorithm to optimize the weights of the Similo algorithm and evalute whether a global combinations of weights can be found. 
We applied the genetic algorithm as follows. 
We begin by fine-tuning the similarity functions for each property, starting from an initial baseline of weights and functions. For each property, we evaluate every possible similarity function, choosing the most effective one. This method is systematically applied to each property in sequence, optimizing them one at a time. 
The selection of property similarity functions might change based on the order in which the properties are optimized. To ensure a robust outcome, we randomly select a property for comparison and conduct multiple rounds of optimization.
In cases where several similarity functions perform well for a property, we employ a brute force approach to determine the best combination from this narrowed selection of functions.
We define a similarity function for each attribute and then apply genetic optimization to determine the optimal set of weights.
We use fixed step values with $0.05$ step size in $[0, 3]$ as possible values for weights. 
Essentially, the algorithm assigns a weight of zero to any attribute that does not enhance overall fitness, thereby excluding it from consideration in the algorithm.
The specific fitness function that evaluates a given weight combination vary based on the optimization goal. More details are provided in \autoref{sec:metrics}.

\subsubsection{\vonsimiloplusplus}
\label{sec:von_similo_pp}

\vonsimiloplusplus is an improved version of \vonsimilo, optimized with the same process as \similoplusplus but with a different objective function: instead of trying to locate the concrete target, an element will also be allowed as a match if it is in the visual or textual overlap or the target is among the top ten highest ranked elements. 

\subsubsection{\hybridsimilo}
The \hybridsimilo approach aims to overcome the limitations of VON Similo mentioned in \textbf{T4}. As discussed, approaches using visual overlap can only identify all elements in the visual overlap with the same accuracy, but not the exact element described by the locator. This can lead to problems in certain testing scenarios, as the element selected in the visual overlap might not be the exact target. On the other hand, Similo can identify the one single concrete element with high accuracy. A preliminary benchmark showed that the basic \vonsimilo algorithm could directly identify the target element in 85.5\% of cases and rank it among the top five in 95.5\% of cases. At the same time, Similo was able to identify the target element in 88\% of cases directly and in 94\% among the top five. While the basic \vonsimilo algorithm is superior in selecting all the elements in the visual overlap, it is inferior in selecting the exact target element.

The idea is to overcome the limitations of \vonsimilo by combining both algorithms, leveraging the strengths of both. We pre-select the elements in the visual overlap using \vonsimilo and then identify the concrete target among them using Similo. In theory, the visual overlap identified by VON Similo should contain elements which differ in their tags, attributes and properties, as the contribute to the formation of a visually cohesive unit. Similo should be able to correctly select the concrete target from the this pre-selection made by \vonsimilo. As the algorithm combines Similo and VON Similo, we named it \hybridsimilo.


\subsection{Metrics}\label{sec:metrics}

Across all two replicated papers, the authors used different evaluation metrics. The metrics proposed in the existing papers are:

\head{Metric 1 $\cdot$ Similo}\label{sec:metric1}
For each of the pairs, the algorithm is tasked to find the the $C'$ among all candidates. If the selected candidate is equal to $C'$ or a direct parent or child in the DOM structure of $C'$ the selection is considered a match. 

\head{Metric 2 $\cdot$ VON Similo}
For the evaluation the authors compared all pairs of original and updated elements by calculating the normalized Similo and VON Similo score between the pair and classifying the pair a match if the score exceeded a certain threshold. They found that for VON Similo a threshold of 0.4 was optimal and for Similo a threshold of 0.28. 

In our replication, we consider three additional metrics. 



\head{Metric 3 $\cdot$ } \textbf{Visual or Textual Overlap.}
Similar to metric 1, the algorithm is tasked to find the correct element among all candidates on an updated version of a websites given the target. A localization is deemed correct if the visual overlap or textual overlap of the located element contains the concrete target. This metrics deems significantly more localization's as correct compared to the more restrictive metric used in Similo. This metric was also used in the successor LLM VON Similo.

\head{Metric 4 $\cdot$ Exact Match}
This new metric determines a located candidate a match if the located element without overlapping elements is the exact target. This is the most restrictive metric but also the most accurate one. A high score indicates that the algorithm is able to correctly identify the target element and in a testing scenario all possible use cases of a locator can be handled (i.e. clicking on an element, entering text, comparing properties).

\head{Metric 5 $\cdot$ Locator Changed - Exact Match}
This metric focuses on an algorithms performance specifically for elements where traditional unique locators (ID, XPath, or ID-XPath) have failed due to website updates. It is similar to  Metric~4, which considers all element pairs, but Metric~5 focuses exclusively on the subset of elements that represent the cases where a real test suite would fail and manual locator repair would be needed. 
By focusing on these cases, we can evaluate the effectiveness of different algorithms specifically in situations where conventional locator strategies fail between updates to the website. This metric provides a more focused assessment of the algorithm's capability to address the main challenge in web test maintenance, correctly recovering from broken locators without manual intervention.

\head{Metric 6 $\cdot$ Fitness} 
In the extended benchmark, we collected detailed information about the variations between different versions, specifically focusing on the elements and their locators. We categorized the element changes into three types, namely \textit{No change}, \textit{Minor change}, and \textit{Major change}.
\textit{No change} refers to elements that retain identical attributes across versions. \textit{Minor change} includes elements that maintain the same tag, text, and attributes, with their locations shifting by no more than 10 pixels in any direction and dimensions changing by no more than 5 pixels, including modifications in the CSS style. We classify as \textit{Major change} all other evolution patterns.

Additionally, the elements were categorized based on their locators into three groups, namely \textit{All locators work}, \textit{Absolute XPath does not work}, \textit{No locators work}. We then assigned a localization score to each element pair based on these categories (we report the actual scores in our replication package). Higher scores were given for more significant changes. The overall fitness score for each pair is the aggregate of their localization scores, rewarding the full score if the exact element was identified and one-quarter of the score if there was only partial overlap.


    


    

In this replication study, we evaluate the two original algorithms (i.e., \similo and \vonsimilo), along with the our new extensions, using all evaluation metrics.

\subsection{Procedure}


Concerning RQ\textsubscript{0}, we executed \similo on the original \similo benchmark, \similo and \vonsimilo on the original \vonsimilo benchmark. 

Concerning RQ\textsubscript{1}, we executed the different \similo and \vonsimilo from each paper on the \similo and \vonsimilo benchmarks as well as our extended benchmark. For all combinations we captured all metrics M1-M6, if feasible. 

Concerning RQ\textsubscript{2}, we employed the optimization process described in \autoref{sec:similopp} to the \similo and \vonsimilo algorithm on different benchmarks and using different fitness metrics. Specifically we optimized Similo on the LLM VON Similo benchmark, as it is the most recent one, for metric M3 and M4 and VON Similo for M3, to provide comparability with the other algorithms. On the extended benchmark we optimized Similo for M6. 

Concerning RQ\textsubscript{3}, we utilized the optimized algorithms from RQ\textsubscript{2}, specifically \vonsimilo optimized on the \similo benchmark to improve M3, as well as \similo optimized on the \similo benchmark to improve M4. We then combined these to first select a set of candidates with \vonsimilo and then determine the exact match among those candidates with \similo.

Overall, our experiment includes nine algorithm configurations under test, four original ones by selecting \similo and \vonsimilo from the original papers as well as seven new ones. As our evaluation set comprises 10,376 element pairs overall, we ran 280,233 localization attempts in our replication and extended study. 

\section{Results}


\autoref{tab:bechmark_results_small} shows the results for RQ\textsubscript{0-1-3}. 
The columns show the results for different metrics with the number of localization complying with the metric first and the percentage of the total number of localizations in brackets. The metric in parenthesis indicates the benchmark the algorithm was optimized on as well as the metrics being optimized by the genetic algorithm. I.e. (LLM, M4) means that the algorithm was optimized on the LLM VON Similo benchmark and the metric M4 was used as a fitness function.
\hybridsimilo is not applicable to M2, because the process of pre-selection does not work with the study setup, where a concrete element pair is given and the score needs be calculated. The columns are each for a specific variation of Similo. For the \similo algorithms taken from the original papers (\similo and \vonsimilo) are followed by the specific paper they are taken from in brackets. For the extended algorithms (\similoplusplus, \vonsimiloplusplus, \hybridsimilo) the content of the brackets indicates on which benchmark the algorithm was optimized and what metric was used as a fitness function. Results are reported separately for each considered benchmark. For metrics M2 and M6 the underlying numbers, i.e., the number of elements correctly classified by the threshold and the exact fitness have no informative value, why we only report the percentage.

\begin{table}[t]
\caption{RQ\textsubscript{0-1-3}: Results for all algorithms across all benchmarks and evaluation metrics (best results are highlighted in bold).}
\label{tab:bechmark_results_small}

\setlength{\tabcolsep}{1pt}
\renewcommand{\arraystretch}{1.1}

\centering
\scriptsize

\begin{adjustbox}{angle=0}

\resizebox{\textwidth}{!}{
\begin{tabular}{@{}lllllll@{}}
    
\toprule

& M1 & M2  & M3 & M4 & M5 & M6 \\
& (Similo) & (VON)  & (LLM) & (Exact) & (LC Exact) & (Fitness) \\

\midrule
\multicolumn{7}{c}{Benchmark used for the Similo Paper (809 Element Pairs, 510 with broken locators)} \\
\midrule

Similo (Similo)        &    720 (88.9\%) &    83.7\% &    720 (88.9\%) &    701 (86.6\%) &    406 (79.6\%) &    83.1\% \\
Similo (VON)           &    727 (89.8\%) &    85.6\% &    725 (89.6\%) &    705 (87.1\%) &    410 (80.4\%) &    83.8\% \\
VON Similo (VON)       &    722 (89.2\%) &    92.8\% &    722 (89.2\%) &    627 (77.5\%) &    353 (69.2\%) &    76.5\% \\
Similo++ (LLM, M4)     &    758 (93.7\%) &    75.6\% &    754 (93.2\%) &    734 (90.7\%) &    436 (85.5\%) &    88.4\% \\
Similo++ (LLM, M3)     &    760 (93.9\%) &    83.6\% &    758 (93.7\%) &    715 (88.4\%) &    417 (81.7\%) &    85.5\% \\
VON Similo++ (LLM, M3) &    728 (89.9\%) &    89.1\% &\bf 765 (94.6\%) &    540 (66.7\%) &    299 (58.6\%) &    65.7\% \\
\hybridsimilo (LLM, M4)&\bf 766 (94.7\%) &    N/A    &    764 (94.4\%) &\bf 742 (91.7\%) &\bf 443 (86.8\%) &\bf 89.7\% \\
Similo++ (Ext, M6)     &    743 (91.8\%) &    73.2\% &    737 (91.1\%) &    713 (88.1\%) &    418 (81.9\%) &    85.4\% \\
\hybridsimilo (Ext, M6)&    749 (92.6\%) &    N/A    &    745 (92.1\%) &    719 (88.8\%) &    424 (83.1\%) &    86.4\% \\

\midrule
\multicolumn{7}{c}{Benchmark used for the VON Similo Paper (441 Element Pairs, 272 with broken locators)} \\
\midrule

Similo (Similo)        &    395 (89.6\%) &    83.4\% &    395 (89.6\%) &    384 (87.1\%) &    217 (79.8\%) &    82.4\% \\
Similo (VON)           &    399 (90.5\%) &    85.2\% &    397 (90.0\%) &    385 (87.3\%) &    218 (80.1\%) &    82.8\% \\
VON Similo (VON)       &    399 (90.5\%) &    93.6\% &    397 (90.0\%) &    323 (73.2\%) &    173 (63.6\%) &    71.9\% \\
Similo++ (LLM, M4)     &    415 (94.1\%) &    76.1\% &    410 (92.9\%) &    398 (90.2\%) &    229 (84.2\%) &    86.1\% \\
Similo++ (LLM, M3)     &    416 (94.3\%) &    84.7\% &    412 (93.4\%) &    381 (86.4\%) &    212 (77.9\%) &    81.4\% \\
VON Similo++ (LLM, M3) &    403 (91.4\%) &    92.3\% &\bf 415 (94.1\%) &    254 (57.6\%) &    126 (46.3\%) &    54.5\% \\
\hybridsimilo (LLM, M4)&\bf 417 (94.5\%) &    N/A    &    414 (93.8\%) &\bf 402 (91.1\%) &\bf 233 (85.6\%) &\bf 88.1\% \\
Similo++ (Ext, M6)     &    406 (92.1\%) &    74.1\% &    399 (90.5\%) &    385 (87.3\%) &    219 (80.5\%) &    83.3\% \\
\hybridsimilo (Ext, M6)&    407 (92.3\%) &    N/A    &    402 (91.1\%) &    386 (87.5\%) &    220 (80.8\%) &    83.7\% \\

\midrule
\multicolumn{7}{c}{Benchmark used for the LLM VON Similo Paper (803 Element Pairs, 500 with broken locators)} \\
\midrule

Similo (Similo)        &    722 (89.9\%) &    84.2\% &    722 (89.9\%) &    703 (87.5\%) &    404 (80.8\%) &    84.2\% \\
Similo (VON)           &    729 (90.8\%) &    86.2\% &    727 (90.5\%) &    707 (88.0\%) &    408 (81.6\%) &    84.9\% \\
VON Similo (VON)       &    724 (90.1\%) &    92.9\% &    724 (90.1\%) &    629 (78.3\%) &    351 (70.2\%) &    77.5\% \\
Similo++ (LLM, M4)     &    760 (94.6\%) &    75.7\% &    757 (94.2\%) &    737 (91.8\%) &    435 (87.00\%)&    89.7\% \\
Similo++ (LLM, M3)     &    762 (94.9\%) &    84.1\% &    761 (94.7\%) &    718 (89.4\%) &    416 (83.2\%) &    86.8\% \\
VON Similo++ (LLM, M3) &    731 (91.0\%) &    89.1\% &\bf 768 (95.6\%) &    544 (67.7\%) &    298 (59.6\%) &    66.8\% \\
\hybridsimilo (LLM, M4)&\bf 768 (95.6\%) &    N/A    &    767 (95.5\%) &\bf 745 (92.8\%) &\bf 442 (88.4\%) &\bf 91.1\% \\
Similo++ (Ext, M6)     &    745 (92.8\%) &    73.6\% &    740 (92.1\%) &    716 (89.1\%) &    417 (83.4\%) &    86.7\% \\
\hybridsimilo (Ext, M6)&    751 (93.5\%) &    N/A    &    748 (93.1\%) &    722 (89.9\%) &    423 (84.6\%) &    87.7\% \\

\midrule
\multicolumn{7}{c}{Extended Benchmark (10376 Element Pairs, 2012 with broken locators)} \\
\midrule

Similo (Similo)        &    10276 (99.0\%) &    89.1\% &    10275 (99.0\%) &    10274 (99.0\%) &    1926 (95.8\%) & 97.6\% \\
Similo (VON)           &    10286 (99.1\%) &\bf 91.4\% &    10285 (99.1\%) &    10282 (99.1\%) &    1934 (96.1\%) & 97.8\% \\
VON Similo (VON)       &    10168 (97.9\%) &    90.3\% &    10270 (98.9\%) &     9448 (91.0\%) &    1689 (83.9\%) & 90.6\% \\
Similo++ (LLM, M4)     &    10322 (99.5\%) &    79.1\% &    10324 (99.5\%) &    10321 (99.4\%) &    1957 (97.3\%) & 98.6\% \\
Similo++ (LLM, M3)     &    10328 (99.5\%) &    85.1\% &    10322 (99.5\%) &    10320 (99.4\%) &    1958 (97.4\%) & 98.6\% \\
VON Similo++ (LLM, M3) &     9779 (94.2\%) &    90.8\% &    10308 (99.3\%) &     8,000 (77.1\%)&    1415 (70.4\%) & 78.1\% \\
\hybridsimilo (LLM, M4)&    10311 (99.4\%) &    N/A     &   10307 (99.3\%) &    10304 (99.3\%) &    1940 (96.4\%) &    98.1\% \\
Similo++ (Ext, M6)     &\bf 10356 (99.8\%) &    79.7\% &\bf 10356 (99.8\%) &\bf 10352 (99.7\%) &\bf 1988 (98.8\%) &\bf 99.4\% \\
\hybridsimilo (Ext, M6)&\bf 10356 (99.8\%) &    N/A     &\bf10356 (99.8\%) &\bf 10352 (99.7\%) &\bf 1988 (98.8\%) &\bf 99.4\% \\

\bottomrule

\end{tabular}}

\end{adjustbox}

\end{table}

\subsection{Replication (RQ\textsubscript{0})}

For Similo, the replicated algorithm and benchmark found 88.99\% elements, where the original paper was able to locate 88.64\% of them. For VON Similo used in the LLM VON Similo paper, our replication found 91.65\% of elements, where the original paper found 91.29\% of elements. These minor differences were expected, as we reloaded the benchmarks from the Web Archive and differences in browser versions and window size can alter coordinates, shapes and areas as well as neighboring text used in the algorithm. 
When replicating Similo and \vonsimilo algorithms on the \vonsimilo benchmark, we also found challenges associated with the use of random non-fitting element pairs in the original benchmark, as the specific elements can change the results. 


\begin{tcolorbox}[boxrule=0pt,sharp corners,boxsep=2pt,left=2pt,right=2pt,top=2.5pt,bottom=2pt]
\textbf{RQ\textsubscript{0} (replication)}: \textit{
Our reproduced results closely match the original outcomes for both Similo and VON Similo, indicating that our replication was successful.
}
\end{tcolorbox}

\subsection{Comparison (RQ\textsubscript{1})}

Our results show that our assumption that VON Similo performs worse than Similo in locating a concrete element and not just its visual overlap is correct. In our experiments, VON Similo under performs Similo on all benchmark, except the VON Similo benchmark, on the metric M3. 
This reinforces our initial hypothesis that Similo excels not only in identifying exact elements but also in recognizing their visual overlaps.
VON Similo surpasses Similo in scenarios involving broader overlap criteria, such as M3 (visual or textual overlaps). This suggests that while Similo tends to either rank the correct targets very high, VON Similo consistently ranks them high, albeit not in the top position.

\begin{tcolorbox}[boxrule=0pt,sharp corners,boxsep=2pt,left=2pt,right=2pt,top=2.5pt,bottom=2pt]
\textbf{RQ\textsubscript{1} (comparison)}: \textit{
Similo is more effective than VON Similo at retrieving the exact target element, while VON Similo performs better when broader visual or textual overlap is sufficient.
}
\end{tcolorbox}

\subsection{Improvements (RQ\textsubscript{2})}

\begin{table}[t]

\caption{RQ\textsubscript{2}: Optimized weights and similarity functions for \similo and \vonsimilo.}
\label{tab:found_properties}

\setlength{\tabcolsep}{2pt}
\renewcommand{\arraystretch}{1.1}
\centering
\scriptsize

\resizebox{\textwidth}{!}{
\begin{tabular}{@{}lllllllll@{}}

\toprule

        & \multicolumn{2}{l}{Similo (Ext M6)} & \multicolumn{2}{l}{VON Similo (Sim. M3)} & \multicolumn{2}{l}{Similo (Sim. M4)} & \multicolumn{2}{l}{Similo (Sim. M3)} \\

        \midrule

        \parbox[t]{1.3cm}{Property} & \parbox[t]{0.4cm}{Wei.} & \parbox[t]{1.55cm}{Sim.} & \parbox[t]{0.4cm}{Wei.} & \parbox[t]{1.55cm}{Sim.} & \parbox[t]{0.4cm}{Wei.} & \parbox[t]{1.55cm}{Sim.}  & \parbox[t]{0.4cm}{Wei.} & \parbox[t]{1.55cm}{Sim.} \\
        \midrule

        Tag            & 0.80  &  Jaccard          & 1.25 & Levenshtein       & 2.35 & Jaro Winkler  & 0.80  & Levenshtein \\ 

        Class          & -    &  -                  & 0.65 & Jaro Winkler      & 1.00 & String Set     & 1.1  & Levenshtein \\ 

        Name           & 2.85 &  Levenshtein         & 1.80  & Equality       & 2.90  & Equality        & 1.70  & Equality \\ 

        ID             & 0.50 &  Levenshtein         & 2.50  & Jaccard       & 2.70  & Levenshtein  & 2.85 & Levenshtein \\ 

        HRef           & 0.95 &  Equality           & 0.80  & Levenshtein & 0.30  & Levenshtein  & 2.85 & Levenshtein  \\ 

        Alt            & 1.85 &  Equality           & 0.10 & Levenshtein & 1.95 & Levenshtein  & 0.60  & Levenshtein  \\ 

        Type           & 2.75 &  Equality          & 2.85 & Equality       & 1.10  & Equality        & 2.45 & Equality \\ 

        Aria-Label     & 0.90  &  Jaccard             & 2.35 & Levenshtein     & 2.95 & Equality        & 1.40  & Equality\\ 

        Abs. XPath & 0.10 &  Jaccard             & 1.05 & Levenshtein & 0.50 & Equality        & 0.05 & Equality \\ 

        ID-XPath       & 0.45 &  Levenshtein      & 0.75 & Equality & 0.50 & Levenshtein  & 1.25 & Levenshtein \\ 

        Is Button      & -    & -                    & 2.85 & Equality      & -    & -            & 0.10 & Equality    \\ 

        Location       & 1.20  & Medium Decay           & 2.00  & Small Decay      & 2.00  & Manhattan    & 2.15 & Linear \\ 

        Dimension      & 0.35 &  Area                   & 0.95 & Area        & 1.30  & Area         & 1.85 & Area \\ 

        Visible Text   & 2.80 &  Levenshtein      & 2.50  & Levenshtein & 2.95 & Levenshtein  & 2.55 & Levenshtein \\ 

        Neighbor Text & 1.45 &  String Set         & 2.30  & Levenshtein   & 1.00  & Levenshtein  & 1.70  & String Set  \\ 

        Attributes     & 1.80  & Intersect Value        & 1.00  & Intersect Value  & 2.20  & Intersect Value   & 2.50 & Intersect Value\\

    \bottomrule
    \end{tabular}
    }
\end{table}

\autoref{tab:found_properties} shows the concrete values we found for each of the property and optimized algorithm. The header shows the algorithm we optimized, the benchmark it was done on and the metric we used as the optimization objective. 

We found that different sets of properties, similarity functions and weights can significantly improve the capabilities of \similo and \vonsimilo. On the LLM VON Similo benchmark, the optimization of Similo improved the M4 (exact match) metric from 87.54\% to 91.78\% and the M3 (overlap visual and textual) from 91.65\% to 95.64\%. While the original algorithms scored very high, in our study we show that there is room for improvement by optimizing the chosen properties, weights and similarity functions. 
Because this optimization used a very broad set of websites, with different specifics due to different web frameworks and different web design, we assume that the optimization would be even more effective for a smaller, more specific set of websites or for a specific web framework.

Some properties like class or is button are often ranked low or excluded entirely from the algorithm. Other properties such as the name, type, aria-label, location, visible text, neighbor text and attributes have high weights no matter the metric, indicating that they are important properties for Similo. 
It is important to note that the optimization process is random and might return different local optima, depending on the initial values. This explains outliers like the sudden high weight of ``is button'' for VON Similo (Sim. M3). 

\begin{tcolorbox}[boxrule=0pt,sharp corners,boxsep=2pt,left=2pt,right=2pt,top=2.5pt,bottom=2pt]
\textbf{RQ\textsubscript{2} (improvements)}: \textit{
Optimizing the property set, similarity functions, and weights yields clear improvements for both Similo and VON Similo, showing that their default configurations are not optimal. Key attributes (e.g., name, type, aria-label, text, location) consistently receive high importance, while others contribute little. Gains are likely even larger when optimizing for specific websites or frameworks.
}
\end{tcolorbox}

\subsection{Hybrid (RQ\textsubscript{3})}

Combining \similo and \vonsimilo can improve the accuracy of the locator re-localization, but only in specific configurations. On the extended benchmark, \similo performs better than \vonsimilo in identifying the visual overlaps. Therefore, we utilize \vonsimilo to find the the top ten highest ranking elements, as it is the only task where \vonsimilo outperforms \similo. Nonetheless, the resulting \hybridsimilo algorithm, which utilizes \vonsimiloplusplus to pre-select ten candidates and \similoplusplus to locate the concrete target, \hybridsimilo performs similar to \similoplusplus for all metrics. 

On the original benchmarks, specifically the LLM VON Similo benchmark, \vonsimiloplusplus proves to be the best algorithm at selecting the visual or textual overlap (M3). Based on the pre-selected overlap by \vonsimiloplusplus, we then used \similoplusplus to find the concrete element. This specific \hybridsimilo version outperforms all other algorithms on the M1 and M4-M6 metric on all original benchmarks. Particularly, it is able to locate 95.51\% of elements on the LLM VON Similo benchmark under the M3 metric, slightly outperforming LLM VON Similo (95.0\%)~\cite{llmvonsimilo}.
These results suggest that \vonsimilo is effective in identifying an initial selection when there is a significant difference between versions. This selection can then be refined by \similo. However, \vonsimilo does not provide additional advantages in case of minor updates between versions.


\textcolor{black}{To ensure that we did not overfit on the benchmark data, we conducted temporal cross-validation on the extended benchmark, using the last $n$ snapshots over all sites as test data and training on all earlier snapshots in five folds. This setup best reflects the common use case of the algorithm, where it is deployed in a functioning test suite and later tasked with locating changed elements on the same website. This approach ensures that our results translate to common testing scenarios. The results show that the performance of the optimized versions on the extended benchmark differs by only 0.1\% on all metrics when using cross-validation. \similoplusplus and \hybridsimilo again performing similarly.}

\begin{tcolorbox}[boxrule=0pt,sharp corners,boxsep=2pt,left=2pt,right=2pt,top=2.5pt,bottom=2pt]
\textbf{RQ\textsubscript{3} (hybrid)}: \textit{
Hybridizing Similo and VON Similo offers benefits only in specific cases. VON Similo is useful for generating an initial shortlist when versions differ substantially, and Similo can then refine this selection to identify the exact element. This hybrid approach improves accuracy on benchmarks with large visual or structural changes, but provides little advantage when updates are minor. Temporal cross-validation confirms that these effects generalize, with hybrid and optimized Similo performing similarly.
}
\end{tcolorbox}


\subsection{Threats to Validity}\label{sec:ttv}

\subsubsection{Internal Validity}

We compared all variants of the replicated and extended algorithms under identical experimental settings and on the same evaluation sets, which was not the case in the original studies. 
The main threat to internal validity concerns our implementation of the original algorithms and evaluation scripts, which we tested thoroughly. Our replication of the original results confirms the correctness of our replication efforts.



\subsubsection{External Validity}

The limited number of websites in our evaluation poses a threat in terms of generalizability of our results to other web apps. We assume that the datasets comprising the most popular websites provides a realistic representation of how websites change over time, as seen in a continuous integration environment. \textcolor{black}{However, different web frameworks and libraries that are not represented in our dataset could yield different results, and systematically evaluating them is left for future research.} 
Furthermore, only the static front pages of websites were used for the dataset. Front pages typically consist of links, headers, images, and menu items and may not represent the diversity of elements found in other parts of a web application. This selection bias could affect the algorithm's effectiveness on pages different than front pages, where elements such as selects, tables, and table items appear more frequently. The final algorithm, with its optimized weights and similarity functions, is tailored to this dataset. Despite using cross-validation to prevent overfitting, there is a risk that the results are overly optimized for the dataset, potentially affecting the algorithm's performance on different websites. 
The data was scraped from the Wayback Machine, which may not always capture a website's complete or accurate representation. 
Additionally, the snapshots are scraped across different browsers in different countries and further pre-processed before saving and rendering them, introducing further inaccuracies into the dataset. 

All these factors could lead the optimized algorithm to perform differently in sanitized testing environments. 




\section{Discussion}\label{sec:discussion}

\head{Discussing Main Differences}
\textcolor{black}{Our results confirm that our Similo and VON Similo replications closely match the original studies and Similo consistently outperforms VON Similo on localizing a concrete element, while VON Similo is only superior on broader overlap metrics (M2/M3). When optimizing properties, similarity functions, and weights yields small but meaningful gains at high baselines (e.g., Similo++ and VON Similo++ improve exact-match accuracy by up to 5.6 percentage points and achieve up to 71\% relative improvement on broken-locator cases) and a hybrid variant using VON Similo++ for candidate pre-selection and Similo++ for final ranking delivers the best overall performance on the original benchmarks and remains competitive on the extended dataset. This indicates that VON Similo is most useful as a overlap pre-filter under larger changes.}

Column M2 (i.e. the metric used for the VON Similo algorithm) differs significantly from the other metrics, as it is the only metric on which VON Similo outperforms all other algorithms. 
This difference is primarily due to variations in the study setup compared to those used for Similo and LLM VON Similo. 
The VON Similo reported than 94.1\% of element pairs were accurately classified as either matching or non-matching. Given that in case of a localization in a test case the target $T$ will be compared with all $C$, we would require $|C| - 1$ correct classifications as non-matching (sensitivity being 0.97) and one as matching (recall being 0.922). 
With approximately 800 elements in $C$ to compare with $T$, the probability of correctly identifying the target in the new version is calculated as $0.97^{799} \cdot 0.92$, which approximates to $0$.


Our results show that VON Similo is not substantially better than Similo, even for identifying the visual overlap. Based on our efforts in hybridizing \similo and \vonsimilo,  we suggest that these algorithms should always be used jointly, as the concrete selection with Similo is expected to improve the locator detection accuracy.




\head{Consequences for use in practical web testing} 
The successfully replicated results show that the proposed algorithms are effective at identifying web elements, even when their standard locators (\id, \xpath, \idxpath) are not working. The original Similo algorithm is able to correclty identify 95.8\% of elements whose original locators are not working anymore. This means that only 4 out of 100 locator breakages need to be manually repaired by a developer, significantly reducing the cost of maintaining a web application. The optimized version of Similo even reduces this number to 1 out of 100. For example, other prominent locator strategies like Tag + Text only achieve a success rate of around 82\%. With the help of the implemented library, developers can directly incorporate the algorithms into their testing process to save time and effort.


Empirical maintenance studies put concrete numbers on the cost of manual locator repair. In an industrial case study involving four real-world Selenium suites, Leotta et al.\ report that repairing a single release after locator breakage requires 0.60 h when ID locators are used and 3.05 h when XPath locators are used \cite{Leotta2013}. At the current U.S. market rate for a test-automation engineer (average \$90 k p.a.\ \cite{IndeedQAEngineerSalary2025}, equivalent to \$43.5 h$^{-1}$), this corresponds to \$26 and \$133 per fix. Assuming a medium-sized organization with 500 test suites $\times$ 10 locators (5,000 locators overall) releasing weekly, and observing that 26\% of XPath locators break (1,300 failures) and in 19\% neither ID nor XPath work (950 failures), maintaining pure, XPath suites would cost approximately \$8.6 million annually, compared to \$1.2 million for ID-based suites. These findings echo Accenture's independently reported \$50–\$120 million annual spend on GUI-test maintenance \cite{Grechanik09}. 

With the original Similo algorithm—automatically recovering 96\% of those 950-1300 weekly dual-break failures—manual repairs drop to just 38-52 per release (\$988-\$6,916). The optimized Similo variant (99\% recovery) further reduces these to around 10-13 repairs (\$260-\$1,729) per release. Over a 50-release year, this cuts annual maintenance from \$1.2-\$8.6 million (no automated healing) to about \$49,400-\$345,800 with Similo or only \$13,000-\$86,450 with optimized Similo, a 96–99 \% reduction in spend. Even marginal gains in locator-healing accuracy thus translate into five-figures annual savings in good maintained test suites or six-figure annual savings in those using fragile locators. This underlines the clear industrial value of further algorithmic improvements.

\head{Impact of long and short inter-version time intervals on performance and benchmark}
At first glance, it may appear that improved methods offer negligible benefits for benchmarks with shorter inter-version intervals since baseline algorithms already perform strongly. However, considering improvements relative to the maximum achievable performance reveals their true significance. For instance, in long-term intervals, performance improved from $86.6\%$ to $91.7\%$, representing $38\%$ of the possible remaining margin towards $100\%$. For short-term intervals, the baseline already achieved $99.0\%$, and improved to $99.7\%$, again representing a substantial $70\%$ of remaining possible improvement. Furthermore, focusing specifically on elements with changed locators, performance improved from $95.8\%$ to $98.8\%$, covering an even larger relative improvement of $71\%$. These results underline that even minor absolute improvements at high baseline levels can translate into significant practical gains, particularly when addressing more challenging locator updates.
 \textcolor{black}{Additionally, we can see that while the Similo algorithms perform better on short-term changes, repairing locators with 98.8\% accuracy, their performance on long-term changes of 1–5 years is significantly lower at 88.4\%, though still better than baseline approaches. This underlines the need for frequent updates to the attribute set of a locator, as done in our implementation.}

\head{Error analysis}
Despite optimization efforts, certain web elements remain consistently misidentified. Common reasons for such errors include:
\begin{itemize}
    \item Substantial changes in attributes between versions (e.g., altered text or tags). This can for example happen when a ``Log in'' button changes from a \texttt{<button>} tag to a styled \texttt{<span>}, significantly reducing similarity scores.
    \item Similarity in visual placement outweighing textual cues. For example, a new ``Sign up'' button positioned exactly like the old ``Log in'' button might cause the algorithm to prioritize visual similarities over differing text.
    \item Inability to interpret purely visual elements like icons. When visually similar icons without textual differences swap positions, the algorithm frequently misclassifies them due to lack of textual or DOM-based identifiers.
\end{itemize}
These issues highlight the need to incorporate additional visual or semantic analysis methods into future approaches.

\head{Need for Standardized Benchmarks in E2E Web Testing}
While this paper provides comparability between the different versions of Similo, we still do not assess their performance on actual running web tests, or locators used in practice by existing tests. 
Indeed, the existing literature on E2E web testing utilizes various benchmarks and evaluation metrics, inhibiting comparisons between tools. Creating a suitable benchmark itself is a challenging task that demands considerable time and effort, which may deter the development of new locator algorithms. 

Nevertheless, we advocate the need for the development of a standardized benchmark for evaluating locator generation algorithms, free from the limitations outlined in \autoref{sec:shortcomings}. Ideally, such a benchmark would encompass a broad spectrum of websites, accommodate frequent updates, and include web elements typical in real-world test scenarios. The availability of a well-curated and consolidated  benchmark would enhance the comparability of tools and provide valuable insights into which algorithms are best suited for specific testing environments, potentially leading to the practical application of these locator algorithms if they prove effective against the current state of the art.


\section{Related Work}\label{sec:related}

We already discussed \similo~\cite{similo}, \vonsimilo~\cite{vonsimilo}, and \llmvonsimilo~\cite{llmvonsimilo}, the first two were  included in the empirical comparison conducted in this work. Besides the approaches investigated in this paper, several techniques have been proposed in literature to re-identify changed web elements in E2E web tests~\cite{Nguyen2021GeneratingAS,color,2015-leotta-ICST,2021-leotta-STVR,2014-leotta-WoSAR,2016-Leotta-JSEP,montoto2010automated,https://doi.org/10.1002/smr.2606,10132210,10.1145/3678869.3685684,10.1007/978-3-031-70245-7_23,ricca2025multiyeargreyliteraturereview,10.1007/978-3-031-43703-8_7,9440153,2021-Ricca-SOFSEM}. We describe the main propositions next. Similarly to the algorithms investigated in this paper, COLOR~\cite{color} re-identifies changed web elements utilizing multiple properties of a target web element.
Leotta et al. propose ROBULA+~\cite{2014-leotta-WoSAR} to create human-readable, short, and simple \xpath{} expressions through an iterative refinement process of a generic \xpath{} until the target element is uniquely identified. 
In follow-up work, the authors have extended the original algorithm by incorporating attribute robustness ranking, excluding fragile attributes, and adding textual information for improved stability~\cite{2016-Leotta-JSEP}. In other works, the robust XPath locator problem is formulated as a graph search problem~\cite{7173589,2021-leotta-STVR}. 
Montoto et al.~\cite{montoto2010automated} propose an algorithm for identifying elements in AJAX websites based on \xpath{} expressions enriched with textual and attribute information.

Concerning ensemble-based strategies, Leotta et al.~\cite{2015-leotta-ICST} introduce the novel concept of MultiLocator in which an ensemble of locator-generating algorithms is used to overcome the limitations of individual approaches. The algorithm captures multiple \xpath{}-based locators for each element. It uses them to identify the element by assigning different voting rights or weights to different locators, choosing the one with the highest vote. 
\llmvonsimilo is the latest iteration of the \similo algorithm~\cite{llmvonsimilo}. The algorithm first uses \vonsimilo to rank all elements on the website by their score. It then takes the top 10 visual overlaps, as the target element is most likely among them. The algorithm then utilizes GPT-4~\cite{gpt4} to find the target element among the pre-selected candidates. The algorithm converts the ten elements and the target element into JSON format. It then sends a request to the large language model with the target and the ten candidates, asking it to identify the target's match. The LLM responds with a single number, indicating the target's position in the list of candidate elements. Since \vonsimilo relies on proprietary OpenAI's APIs and uses an unspecified version of GPT-4 with unknown temperature and configuration settings, its results cannot be reliably reproduced in a controlled experimental environment. Therefore, we do not attempt a full replication of this method in this paper.
In a recent work, Coppola et al.~\cite{COPPOLA2025112286} extend the MultiLocator approach with standard a Learning to Rank solution to facilitate the relocalization of web elements.

All related studies compare their proposed methods with the state-of-the-art solutions available at the time of publication. In contrast, our work provides an in-depth analysis specifically focused on the family of \similo algorithms, evaluating different versions on both existing and newly developed benchmarks. Our replication study represents a new contribution as it is the first of its kind. Other than the replication, this work provides extensions of the original algorithms in two directions (i.e., a weight-optimized variant and a hybrid approach), other than a larger benchmark.

\section{Future Work}\label{sec:futurework}

To further improve the algorithms accuracy, we believe that extending the set of properties to include non-DOM-based properties would be necessary. Analyzing the results revealed that icons were often mismatched, as the algorithm cannot differentiate between them as effectively as it can with text. 

Additionally, future work could address the limitations of using snapshots from the Wayback Machine, such as the reliability and completeness of the data. Exploring alternative sources or methods for obtaining website snapshots, such as utilizing open-source applications with complete version histories, could help mitigate these limitations and provide more accurate and reliable data.

We have observed a correlation between property stability and optimal weights. Finding concrete evidence of correlations between property stability, uniqueness, or other metrics for a specific website could enable a self-tuning and optimizing Similo algorithm. This would allow the algorithm to adapt to the website's properties and changes over time, further improving its accuracy and robustness.

\textcolor{black}{While this work has focused on reproducibility and providing a locator library that enables replication and reuse of our approach, an important direction for future work is a systematic evaluation of the library or a variation in industrial settings. Such an evaluation could assess its effectiveness on nested, evolving web applications, the effort required to integrate it into existing CI pipelines, and the extent to which developers adopt and save time in practice.}



\section{Conclusions}\label{sec:conclusions}

This paper replicates two existing studies about web element locator generation algorithms. We discussed the main threats to the validity of the original studies, and re-designed an experimental design to address them. Moreover, our study extends the original studies with a substantially expanded benchmark and configurations. 

While our experiments did not entirely replicate the findings of the original study, our results did align with those reported in the replicated studies. However, our findings did not confirm the previously reported superiority of VON Similo over Similo. The discrepancies observed were associated with variations in the benchmark used. In our replication effort, we harmonized the experimentation by utilizing consistent benchmarks.

Our results provide strong justification for the development and adoption of well-curated and standardized benchmarks of webpages, tests and locators for E2E web testing research. Such initiative and endeavor could significantly foster the development of innovative locator generation algorithms by allowing researchers to focus on devising new algorithms rather than on constructing benchmarks. Consequently, this standardization would enable the consistent evaluation of new proposals using consolidated datasets.


\section{Declarations} 

\subsection{Funding}
This research was funded by the Bavarian Ministry of Economic Affairs, Regional Development and Energy. 

\subsection{Ethical Approval}  
Not applicable.

\subsection{Informed Consent}  
Not applicable.

\subsection{Author Contributions}  

\textbf{Anton Kluge}: conceptualization, methodology, implementation, evaluation, writing, review, editing. \textbf{Andrea Stocco}: conceptualization, methodology, review, editing.

\subsection{Data Availability Statement}  

All our results, the source code, and the library are accessible and can be reproduced~\cite{replication-package}.

\subsection{Conflict of Interest}  
The authors declare no conflict of interest.

\subsection{Clinical Trial Registration}  
Clinical trial number: Not applicable.

\balance
\bibliographystyle{spmpsci}
\bibliography{paper}

\newpage
\appendix   

\section{Extended Benchmark}

\begin{table}[htbp]
    \caption{Distribution of elements over different websites and locator changes.}
    \centering
    \begin{tabular}{@{}p{0.14\textwidth}*{6}{p{\dimexpr0.14\textwidth-2\tabcolsep\relax}}@{}}
    \toprule
    & \multicolumn{3}{c}{Number of elements} & \multicolumn{3}{c}{Locator changes} \\
    \cmidrule(r{4pt}){2-4} \cmidrule(l){5-7}
    & Initial elements & Elements available on all versions & Number of element pairs & Same locators  & Absolute \xpath{} broke & All locators broke \\
    \midrule
    Accuweather & 21 & 16 & 262 & 93 & 49 & 120 \\
    Adobe & 7 & 6 & 93 & 53 & 9 & 31 \\
    Amazon & 42 & 27 & 506 & 247 & 162 & 97 \\
    Apple & 18 & 17 & 264 & 226 & 4 & 34 \\
    Bestbuy & 45 & 20 & 469 & 145 & 174 & 150 \\
    Bing & 10 & 2 & 58 & 40 & 17 & 1 \\
    Discord & 24 & 15 & 269 & 199 & 16 & 54 \\
    Ebay & 45 & 33 & 538 & 260 & 223 & 55 \\
    Espn & 22 & 13 & 257 & 212 & 20 & 25 \\
    Etsy & 36 & 30 & 467 & 244 & 95 & 128 \\
    Facebook & 24 & 8 & 203 & 87 & 59 & 57 \\
    Fidelity & 54 & 34 & 727 & 579 & 142 & 6 \\
    Google & 15 & 14 & 211 & 92 & 93 & 26 \\
    Indeed & 25 & 19 & 330 & 139 & 107 & 84 \\
    Linkedin & 30 & 17 & 285 & 165 & 0 & 120 \\
    Netflix & 16 & 16 & 240 & 137 & 0 & 103 \\
    Office & 41 & 33 & 578 & 290 & 263 & 25 \\
    Paypal & 32 & 21 & 377 & 265 & 28 & 84 \\
    Reddit & 25 & 2 & 243 & 81 & 61 & 101 \\
    Roblox & 31 & 23 & 369 & 222 & 104 & 43 \\
    T-Mobile & 45 & 27 & 464 & 235 & 135 & 94 \\
    Tiktok & 21 & 1 & 45 & 30 & 2 & 13 \\
    Twitter & 28 & 7 & 308 & 248 & 0 & 60 \\
    Usps & 44 & 35 & 558 & 272 & 283 & 3 \\
    Walmart & 49 & 5 & 332 & 99 & 115 & 118 \\
    Weather & 26 & 19 & 311 & 154 & 85 & 72 \\
    Wikipedia & 21 & 19 & 295 & 152 & 61 & 82 \\
    Yahoo & 24 & 12 & 287 & 77 & 141 & 69 \\
    Zillow & 50 & 20 & 336 & 179 & 43 & 114 \\
    Zoom & 62 & 36 & 694 & 449 & 203 & 42 \\
    
    \midrule
    \textbf{Total} & \textbf{933} & \textbf{547} & \textbf{10376} & \textbf{5671} & \textbf{2694} & \textbf{2011} \\
    \bottomrule
    \end{tabular}
    \label{tab:classification_distribution}
\end{table}

\newpage

\section{Full Benchmark}

This table shows all results for all algorithms that we evaluated, even those that were not deemed relevant for the main paper.

\begin{table}[H]
\caption{RQ\textsubscript{0-1-3}: Results for all algorithms across all benchmarks and evaluation metrics (best results are highlighted in bold).}
\label{tab:bechmark_results_large}

\setlength{\tabcolsep}{1pt}
\renewcommand{\arraystretch}{1.1}

\centering
\scriptsize

\begin{adjustbox}{angle=0}

\resizebox{\textwidth}{!}{
\begin{tabular}{@{}llllllllll@{}}
    
\toprule

& M1 & M2  & M3 & M4 & M5 & M6 & M7 & M8 & M9 \\
& (Similo) & (VON)  & (LLM) & (Exact) & (LC Exact) & (LC Close) & (Vis. Over.) & (Top Ten) & (Fitness) \\

\midrule
\multicolumn{10}{c}{Benchmark used for the Similo Paper (809 Element Pairs, 510 with broken locators)} \\
\midrule

Similo (Similo)        &    720 (88.9\%) &    83.7\% &    720 (88.9\%) &    701 (86.6\%) &    406 (79.6\%) &    425 (83.3\%) &    719 (88.8\%) &    762 (94.2\%) &    83.1\% \\

Similo (VON)           &    727 (89.8\%) &    85.6\% &    725 (89.6\%) &    705 (87.1\%) &    410 (80.4\%) &    432 (84.7\%) &    724 (89.5\%) &    764 (94.4\%) &    83.8\% \\
VON Similo (VON)       &    722 (89.2\%) &    92.8\% &    722 (89.2\%) &    627 (77.5\%) &    353 (69.2\%) &    432 (84.7\%) &    704 (87.0\%) &    773 (95.5\%) &    76.5\% \\
VON Similo (LLM)       &    731 (90.3\%) &\bf 93.1\% &    735 (90.8\%) &    616 (76.1\%) &    338 (66.2\%) &    437 (85.7\%) &    691 (85.4\%) &    785 (97.0\%) &    73.9\% \\
Similo++ (LLM, M4)     &    758 (93.7\%) &    75.6\% &    754 (93.2\%) &    734 (90.7\%) &    436 (85.5\%) &    460 (90.2\%) &    753 (93.1\%) &    781 (96.5\%) &    88.4\% \\
Similo++ (LLM, M3)     &    760 (93.9\%) &    83.6\% &    758 (93.7\%) &    715 (88.4\%) &    417 (81.7\%) &    462 (90.6\%) &    740 (91.5\%) &    792 (97.9\%) &    85.5\% \\
VON Similo++ (LLM, M3) &    728 (89.9\%) &    89.1\% &\bf 765 (94.6\%) &    540 (66.7\%) &    299 (58.6\%) &    446 (87.4\%) &    626 (77.4\%) &    792 (97.9\%) &    65.7\% \\
\hybridsimilo (LLM, M4)&\bf 766 (94.7\%) &    N/A     &    764 (94.4\%) &\bf 742 (91.7\%) &\bf 443 (86.8\%) &\bf 467 (91.6\%) &\bf 763 (94.3\%) &    765 (94.6\%) &\bf 89.7\% \\
Similo++ (Ext, M9)     &    743 (91.8\%) &    73.2\% &    737 (91.1\%) &    713 (88.1\%) &    418 (81.9\%) &    447 (87.6\%) &    736 (90.9\%) &    793 (98.0\%) &    85.4\% \\
VON Similo++ (Ext, M8) &    741 (91.6\%) &    83.3\% &    738 (91.2\%) &    635 (78.5\%) &    358 (70.2\%) &    447 (87.6\%) &    718 (88.7\%) &\bf 797 (98.5\%) &    77.7\% \\
\hybridsimilo (Ext, M9)&    749 (92.6\%) &    N/A     &    745 (92.1\%) &    719 (88.8\%) &    424 (83.1\%) &    453 (88.8\%) &    743 (91.8\%) &\bf 797 (98.5\%) &    86.4\% \\

\midrule
\multicolumn{10}{c}{Benchmark used for the VON Similo Paper (441 Element Pairs, 272 with broken locators)} \\
\midrule

Similo (Similo)        &    395 (89.6\%) &    83.4\% &    395 (89.6\%) &    384 (87.1\%) &    217 (79.8\%) &    228 (83.8\%) &    394 (89.3\%) &    421 (95.4\%) &    82.4\% \\
Similo (VON)           &    399 (90.5\%) &    85.2\% &    397 (90.0\%) &    385 (87.3\%) &    218 (80.1\%) &    232 (85.3\%) &    396 (89.8\%) &    422 (95.7\%) &    82.8\% \\
VON Similo (VON)       &    399 (90.5\%) &    93.6\% &    397 (90.0\%) &    323 (73.2\%) &    173 (63.6\%) &    235 (86.4\%) &    379 (85.9\%) &    428 (97.0\%) &    71.9\% \\
VON Similo (LLM)       &    397 (90.0\%) &\bf 94.0\% &    404 (91.6\%) &    308 (69.8\%) &    152 (55.8\%) &    231 (84.9\%) &    363 (82.3\%) &    431 (97.7\%) &    65.5\% \\
Similo++ (LLM, M4)     &    415 (94.1\%) &    76.1\% &    410 (92.9\%) &    398 (90.2\%) &    229 (84.2\%) &    246 (90.4\%) &    409 (92.7\%) &    426 (96.6\%) &    86.1\% \\
Similo++ (LLM, M3)     &    416 (94.3\%) &    84.7\% &    412 (93.4\%) &    381 (86.4\%) &    212 (77.9\%) &    247 (90.8\%) &    394 (89.3\%) &    433 (98.1\%) &    81.4\% \\
VON Similo++ (LLM, M3) &    403 (91.4\%) &    92.3\% &\bf 415 (94.1\%) &    254 (57.6\%) &    126 (46.3\%) &    238 (87.5\%) &    314 (71.2\%) &    433 (98.1\%) &    54.5\% \\
\hybridsimilo (LLM, M4) &\bf 417 (94.5\%) &    N/A     &    414 (93.8\%) &\bf 402 (91.1\%) &\bf 233 (85.6\%) &\bf 248 (91.1\%) &\bf 413 (93.6\%) &    415 (94.1\%) &\bf 88.1\% \\
Similo++ (Ext, M9)     &    406 (92.1\%) &    74.1\% &    399 (90.5\%) &    385 (87.3\%) &    219 (80.5\%) &    239 (87.8\%) &    398 (90.2\%) &    433 (98.1\%) &    83.3\% \\
VON Similo++ (Ext, M8) &    407 (92.3\%) &    86.1\% &    404 (91.6\%) &    323 (73.2\%) &    172 (63.2\%) &    241 (88.6\%) &    384 (87.1\%) &\bf 436 (98.8\%) &    72.1\% \\
\hybridsimilo (Ext, M9) &    407 (92.3\%) &    N/A     &    402 (91.1\%) &    386 (87.5\%) &    220 (80.8\%) &    240 (88.2\%) &    400 (90.7\%) &\bf 436 (98.8\%) &    83.7\% \\

\midrule
\multicolumn{10}{c}{Benchmark used for the LLM VON Similo Paper (803 Element Pairs, 500 with broken locators)} \\
\midrule

Similo (Similo)        &    722 (89.9\%) &    84.2\% &    722 (89.9\%) &    703 (87.5\%) &    404 (80.8\%) &    423 (84.6\%) &    721 (89.8\%) &    758 (94.4\%) &    84.2\% \\
Similo (VON)           &    729 (90.8\%) &    86.2\% &    727 (90.5\%) &    707 (88.0\%) &    408 (81.6\%) &    430 (86.0\%) &    726 (90.4\%) &    759 (94.5\%) &    84.9\% \\
VON Similo (VON)       &    724 (90.1\%) &    92.9\% &    724 (90.1\%) &    629 (78.3\%) &    351 (70.2\%) &    430 (86.0\%) &    706 (87.9\%) &    770 (95.9\%) &    77.5\% \\
VON Similo (LLM)       &    732 (91.1\%) &\bf 93.2\% &    736 (91.6\%) &    618 (76.9\%) &    335 (67.00\%) &    434 (86.8\%) &    692 (86.1\%) &    781 (97.2\%) &    74.8\% \\
Similo++ (LLM, M4)     &    760 (94.6\%) &    75.7\% &    757 (94.2\%) &    737 (91.8\%) &    435 (87.00\%) &    458 (91.6\%) &    756 (94.1\%) &    776 (96.6\%) &    89.7\% \\
Similo++ (LLM, M3)     &    762 (94.9\%) &    84.1\% &    761 (94.7\%) &    718 (89.4\%) &    416 (83.2\%) &    460 (92.0\%) &    743 (92.5\%) &    786 (97.8\%) &    86.8\% \\
VON Similo++ (LLM, M3) &    731 (91.0\%) &    89.1\% &\bf 768 (95.6\%) &    544 (67.7\%) &    298 (59.6\%) &    445 (89.0\%) &    629 (78.3\%) &    789 (98.2\%) &    66.8\% \\
\hybridsimilo (LLM, M4)&\bf 768 (95.6\%) &    N/A     &    767 (95.5\%) &\bf 745 (92.8\%) &\bf 442 (88.4\%) &\bf 465 (93.0\%) &\bf 766 (95.4\%) &    768 (95.6\%) &\bf 91.1\% \\
Similo++ (Ext, M9)     &    745 (92.8\%) &    73.6\% &    740 (92.1\%) &    716 (89.1\%) &    417 (83.4\%) &    445 (89.0\%) &    739 (92.0\%) &    787 (98.00\%) &    86.7\% \\
VON Similo++ (Ext, M8) &    743 (92.5\%) &    83.9\% &    741 (92.2\%) &    638 (79.4\%) &    357 (71.4\%) &    445 (89.0\%) &    721 (89.8\%) &\bf 794 (98.8\%) &    78.9\% \\
\hybridsimilo (Ext, M9)&    751 (93.5\%) &    N/A     &    748 (93.1\%) &    722 (89.9\%) &    423 (84.6\%) &    451 (90.2\%) &    746 (92.9\%) &\bf 794 (98.8\%) &    87.7\% \\

\midrule
\multicolumn{10}{c}{Extended Benchmark (10376 Element Pairs, 2012 with broken locators)} \\
\midrule

Similo (Similo)        &    10276 (99.0\%) &    89.1\% &    10275 (99.0\%) &    10274 (99.0\%) &    1926 (95.8\%) &    1928 (95.8\%) &    10274 (99.0\%) &    10354 (99.8\%) &    97.6\% \\
Similo (VON)           &    10286 (99.1\%) &\bf 91.4\% &    10285 (99.1\%) &    10282 (99.1\%) &    1934 (96.1\%) &    1938 (96.4\%) &    10284 (99.1\%) &    10361 (99.8\%) &    97.8\% \\
VON Similo (VON)       &    10168 (97.9\%) &    90.3\% &    10270 (98.9\%) &     9448 (91.0\%) &    1689 (83.9\%) &    1934 (96.1\%) &    10267 (98.9\%) &    10356 (99.8\%) &    90.6\% \\
VON Similo (LLM)       &    10230 (98.6\%) &    90.4\% &    10291 (99.1\%) &     9494 (91.5\%) &    1681 (83.6\%) &    1939 (96.4\%) &    10210 (98.4\%) &    10357 (99.8\%) &    90.2\% \\
Similo++ (LLM, M4)     &    10322 (99.5\%) &    79.1\% &    10324 (99.5\%) &    10321 (99.4\%) &    1957 (97.3\%) &    1958 (97.4\%) &    10323 (99.5\%) &    10368 (99.9\%) &    98.6\% \\
Similo++ (LLM, M3)     &    10328 (99.5\%) &    85.1\% &    10322 (99.5\%) &    10320 (99.4\%) &    1958 (97.4\%) &    1966 (97.7\%) &    10321 (99.4\%) &    10371 (99.9\%) &    98.6\% \\
VON Similo++ (LLM, M3) &     9779 (94.2\%) &    90.8\% &    10308 (99.3\%) &     8,000 (77.1\%) &    1415 (70.4\%) &    1810 (90.00\%) &     8,705 (83.9\%) &    10368 (99.9\%) &    78.1\% \\
\hybridsimilo (LLM, M4)&    10311 (99.4\%) &    N/A     &    10307 (99.3\%) &    10304 (99.3\%) &    1940 (96.4\%) &    1947 (96.8\%) &    10306 (99.3\%) &    10315 (99.4\%) &    98.1\% \\
Similo++ (Ext, M9)     &\bf 10356 (99.8\%) &    79.7\% &\bf 10356 (99.8\%) &\bf 10352 (99.7\%) &\bf 1988 (98.8\%) &\bf 1991 (99.00\%) &\bf 10355 (99.8\%) &    10374 (99.9\%) &\bf 99.4\% \\
VON Similo++ (Ext, M8) &    10194 (98.2\%) &    90.6\% &    10314 (99.4\%) &     9475 (91.3\%) &    1699 (84.5\%) &    1944 (96.6\%) &    10313 (99.4\%) &\bf 10375 (99.9\%) &    91.1\% \\
\hybridsimilo (Ext, M9)&\bf 10356 (99.8\%) &    N/A     &\bf 10356 (99.8\%) &\bf 10352 (99.7\%) &\bf 1988 (98.8\%) &\bf 1991 (99.00\%) &\bf 10355 (99.8\%) &\bf 10375 (99.9\%) &\bf 99.4\% \\

\bottomrule

\end{tabular}}

\end{adjustbox}

\end{table}

\section{Cross Validation}

We cross validated our results on five temporal folds, training on older snapshots and testing performance on new ones. The results for \similoplusplus and \hybridsimilo for this validation can be found here.

\begin{table}[t]
\small                         
\caption{\similoplusplus on the extended dataset with target metric M6:
Test set performance across 5-fold temporal cross-validation.}
\label{tab:similo_extended_m6_results}

\setlength{\tabcolsep}{6pt}
\renewcommand{\arraystretch}{1.15}
\centering

\begin{tabular}{@{}lcccc@{}}
\toprule
\textbf{Metric} & \textbf{Mean} & \textbf{Std} & \textbf{Min} & \textbf{Max} \\
\midrule
M1 & 99.76\% & $\pm$0.04 & 99.71\% & 99.82\% \\
M2 & 97.44\% & $\pm$1.27 & 95.38\% & 98.49\% \\
M3 & 99.76\% & $\pm$0.04 & 99.71\% & 99.82\% \\
M4 & 99.75\% & $\pm$0.04 & 99.71\% & 99.82\% \\
M5 & 98.72\% & $\pm$0.26 & 98.39\% & 99.11\% \\
M6 & 99.33\% & $\pm$0.12 & 99.20\% & 99.54\% \\
\bottomrule
\end{tabular}
\end{table}

\begin{table}[t]

\caption{\hybridsimilo on the extended dataset with target metric M6: Test set performance across 5-fold temporal cross-validation.}
\label{tab:similo_extended_m6_results}

\setlength{\tabcolsep}{6pt}
\renewcommand{\arraystretch}{1.15}
\centering
\small

\begin{tabular}{@{}lcccc@{}}

\toprule

\textbf{Metric} & \textbf{Mean} & \textbf{Std} & \textbf{Min} & \textbf{Max} \\

\midrule

M1 & 99.67\% & $\pm$0.11 & 99.48\% & 99.77\% \\
M2 & 97.02\% & $\pm$2.28 & 92.58\% & 98.92\% \\
M3 & 99.66\% & $\pm$0.10 & 99.48\% & 99.76\% \\
M4 & 99.64\% & $\pm$0.09 & 99.48\% & 99.73\% \\
M5 & 98.10\% & $\pm$0.59 & 97.07\% & 98.67\% \\
M6 & 99.03\% & $\pm$0.29 & 98.53\% & 99.31\% \\

\bottomrule

\end{tabular}

\end{table}
\newpage

\end{document}